\documentclass[lettersize,journal]{IEEEtran}
\usepackage{amsmath,amsfonts}
\usepackage{algorithmic}
\usepackage{algorithm}
\usepackage{array}
\usepackage{textcomp}
\usepackage{stfloats}
\usepackage{url}
\usepackage{verbatim}
\usepackage{graphicx}
\usepackage{cite}
\usepackage{subfigure}
\usepackage{booktabs}
\usepackage{bbding}
\usepackage{color}
\usepackage{float}
\usepackage{multirow}
\usepackage[ 
colorlinks=True,
linkcolor=blue
]{hyperref}

\usepackage[switch]{lineno}
\definecolor{Blue}{RGB}{70, 130, 180}
\definecolor{Green}{RGB}{102, 205, 170}

\begin{document}

\title{Look\&Listen: Multi-Modal Correlation Learning for Active Speaker Detection and Speech Enhancement}

\author{Junwen Xiong$^{*}$, Yu Zhou$^{*}$,  Peng Zhang$^{*}$,  Lei Xie, Wei Huang, Yufei Zha
	\thanks{Manuscript received January 29, 2022; revised April 24, 2022 and June 21, 2022. 
	This work was supported in part by National Natural Science Foundation
	of China under Grant 61971352 and Grant 61862043, and in part by Natural Science Foundation of Ningbo under Grant 2021J048 and Grant 2021J049, in part by Natural Science Foundation of Jiangxi Province in China under Grant 20204BC J22011. 
	
	\indent Junwen Xiong, Yu Zhou,  and Lei Xie are with School of Computer Science, Northwestern Polytechnical	University, Xi’an, China.
	 Peng Zhang and Yufei Zha are with School of Computer Science, Northwestern PolytechnicalUniversity, Xi’an, China and with Ningbo Institute of Northwestern Polytechnical University, Ningbo, China.
	 Wei Huang is with School of Mathematics and Computer Sciences, Nanchang University, China.
	Correspondence should be addressed to Peng Zhang (e-mail: zh0036ng@nwpu.edu.cn). \\
	\indent	$^*$The first three authors equally contributed to this work.}
}

\markboth{Journal of \LaTeX\ Class Files,~Vol.~14, No.~8, August~2021}%
{Shell \MakeLowercase{\textit{et al.}}: A Sample Article Using IEEEtran.cls for IEEE Journals}

\maketitle
\begin{abstract}
Active speaker detection and speech enhancement have become two increasingly attractive topics in audio-visual scenario understanding. According to their respective characteristics, the scheme of independently designed architecture has been widely used in correspondence to each single task. This may lead to the representation learned by the model being task-specific, and inevitably result in the lack of generalization ability of the feature based on multi-modal modeling. More recent studies have shown that establishing cross-modal relationship between auditory and visual stream is a promising solution for the challenge of audio-visual multi-task learning. Therefore, as a motivation to bridge the multi-modal associations in audio-visual tasks, a unified framework is proposed to achieve target speaker detection and speech enhancement with joint learning of audio-visual modeling in this study.

With the assistance of audio-visual channels of videos in challenging real-world scenarios, the proposed method is able to exploit inherent correlations in both audio and visual signals, which is used to further anticipate and model the temporal audio-visual relationships across spatial-temporal space via a cross-modal conformer. In addition, a plug-and-play multi-modal layer normalization is introduced to alleviate the distribution misalignment of multi-modal features. Based on cross-modal circulant fusion, the proposed model is capable to learned all audio-visual representations in a holistic process. Substantial experiments demonstrate that the correlations between different modalities and the associations among diverse tasks can be learned by the optimized model more effectively. In comparison to other state-of-the-art works, the proposed work  shows a superior performance for active speaker detection and audio-visual speech enhancement on three benchmark datasets, also with a favorable generalization in diverse challenges. Code is available at: \url{https://github.com/Overcautious/ADENet}.

\end{abstract}


\section{INTRODUCTION}
\label{sec:intro}

\begin{figure}[tbp]
	\centering
	\includegraphics[scale=0.36]{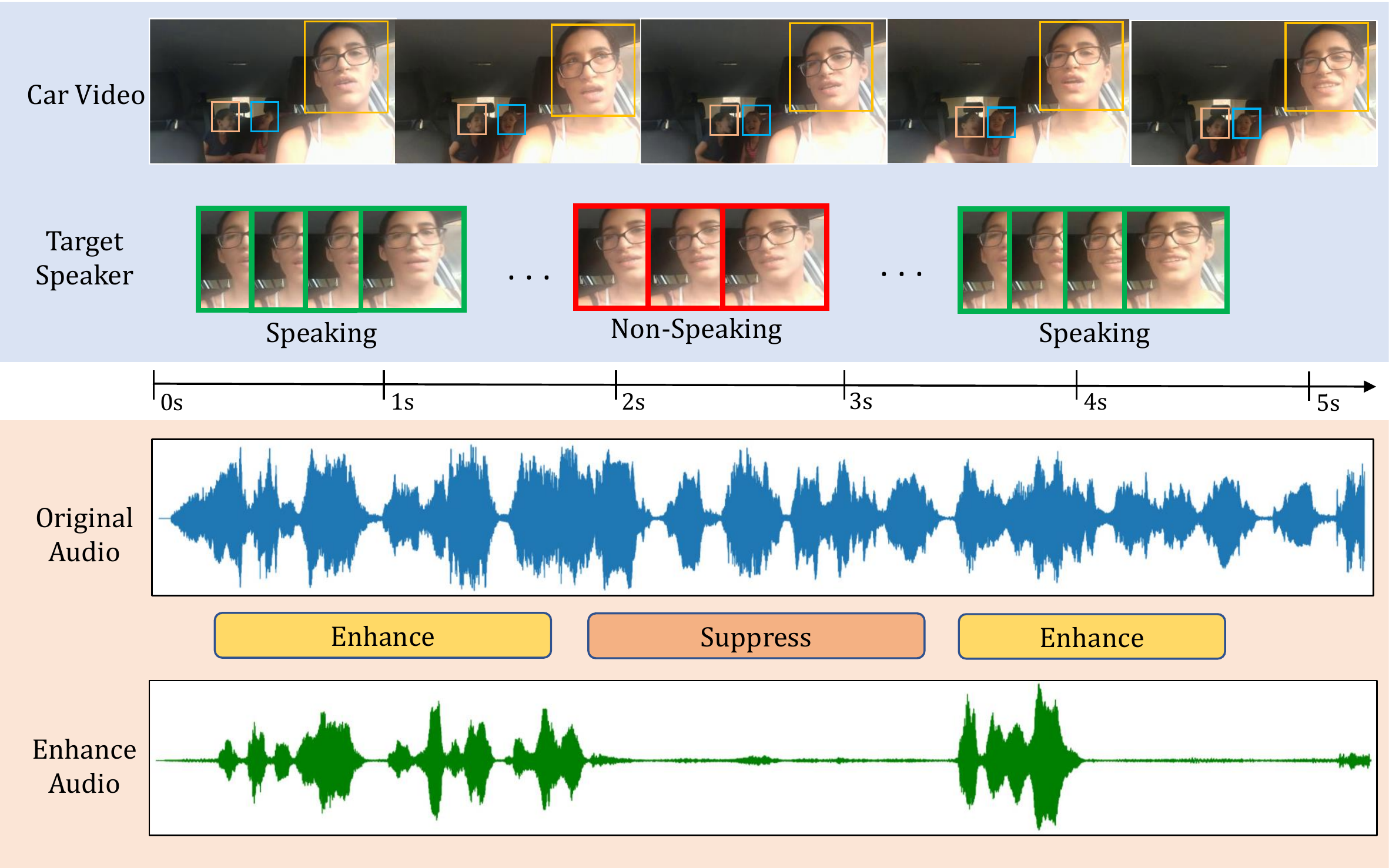}
	\caption{Given a real-world application, our ADENet model can detect the target speaker with the aid of the visual and audio channels, while enhancing the target speaker's speech signal.}
	\label{figure1}
\end{figure}

Vision and audition are two of the most important senses for human brain to perceive the external environments. Even though the auditory and visual information can be very different, the robust perception can be efficiently integrated through multiple senses \cite{ERNST2004162, BURR2006243}. Such a combination originates from the biological phenomena of "sensory cooperation"\cite{bulthoff1988integration}, and exists through our daily life, $e$.$g$., chatting with dining partners, watching a heated presidential debate, meeting with colleagues, and etc. These inherent and pervasive correspondences provide us with the reference to distinguish and correlate different audio-visual events, then contribute to learning diversified associations between visual appearances and their produced sounds. For example, in a driving scenario with multiple speaking and silent objects (shown in Fig. \ref{figure1}), people can easily filter out the silent objects using reliable auditory and visual cues such as accurate lip movements and valid audio signals, as well as identify different speakers  and  enhance the speech of each speaker simultaneously. Unfortunately, highlighting the wanted information from these jointly seen audiovisual scenes can be difficult, especially when perception is frequently hampered by an arbitrary number of input sources, occasional speech overlapping, and visual occlusions.

Existing solutions to the challenges above can be categorized into two directions: $active$ $speaker$ $detection$ (ASD) and $audio$-$visual$ $speech$ $enhancement$ (AVSE). The former seeks to identify active speakers, among a set of possible candidates, by analyzing subtle facial motion patterns and carefully aligning their characteristic speech waveforms \cite{alcazar2020active}. Comparatively, the latter aims to separate the speaker's voice from multi-talkers' simultaneous speeches with the help of visual information. But these two tasks are not explored jointly in the primarily discussed literature.

To accomplish both tasks in a synergetic way, the multisensory integration  of sensory neuroscience \cite{10.1162/jocn.2007.19.12.1964} can be taken into consideration to bridge the interaction between the visual and the auditory perceptions \cite{von2008simulation}. 
Moreover, both ASD and AVSE  require the visual cues like mouth and tongue movements to provide complementary information about speech content.
This means that a joint modeling strategy of two tasks should be more effective to discover active speaking objects and enhance their corresponding voices.


%
%
%



Most previous ASD methods \cite{haider2016active,alcazar2020active, roth2020ava, tao2021someone, kopuklu2021design, zhang2021unicon, barnard2014robust} mainly focus on modeling the audio and visual streams to maximize the performance of target speaker prediction over video segments. Generally, these models consist of two functional sub-networks: an audio-visual encoding network to extract the spatial-temporal audio and visual features, and a relational modeling network to learn the audio-visual inter-modality interaction from the extracted features. For audio-visual encoding network, the 2D and 3D convolutional models are two common types of network architectures being used\cite{he2016deep,tran2015learning}, and for relational modeling network, it is employed to further explore the temporal context information from extracted audio-visual features. Inspired by the progress in speech recognition and natural language processing, current relational modeling networks are designed by recurrent neural networks (RNNs)  \cite{alcazar2020active, roth2020ava, chung2019naver, zhang2019multi} or cross-attention \cite{tao2021someone, li2021ctnet}.


On the one hand, the features extracted from audio-visual encoding network are easy to contaminated by the noisy environment, which leads to a substantial decadence in detection performance. On the other hand, the existing relational modeling networks have taken into consideration the correlations between different modalities, while the intrinsic heterogeneity of audio and visual modalities is yet far from sufficient usage, which causes the distribution misalignment of the two feature domains.


From the aspect of AVSE, all the existing works explicitly assume a local correspondence between the audio and visual streams \cite{gabbay2017visual, ephrat2018looking, afouras2018deep, morrone2019face, ito2021audio, aldeneh2021role}. By analyzing the facial information in concert with the emitted speech, these methods steer the audio enhancement module towards the relevant portions of the sound that ought to be separated out from the full audio track. For example, lip movements in different shapes generate the corresponding sounds, this makes it possible to isolate desired speech of the target speaker based on audio-visual consistency. However, when lip movement is misleading, only depending on lip movement may fail, $e$.$g$., the same mouth shape with different inside tongue movements, which means the different pronunciations. Furthermore, these methods usually focus on simple audiovisual scenarios and fail to characterize some specific sounding objects. Therefore, how to reliably identify visual information of the sounding objects still motivates successive studies like ours.


In this study, a Multi-Modal Correlation Learning for \textbf{A}ctive Speaker \textbf{D}etection and Speech \textbf{E}nhancement (ADENet) is proposed, which is a unified framework enables the audio-visual speech enhancement to be jointly learned together with active speaker detection. In addition, we design a cross-modal conformer as a relationship grabber to encode the temporal context of audio and visual cues from video segments. Instead of directly aggregating audio and visual features, it is necessary to unravel the heterogeneity between them in advance to ensure the distribution alignment of multi-domains. Specifically, a simple and effective layer normalization variant is introduced in the cross-modal conformer to alleviate the problem of distribution misalignment.

\begin{figure}[tbp]
	\centering
	\includegraphics[scale=0.36]{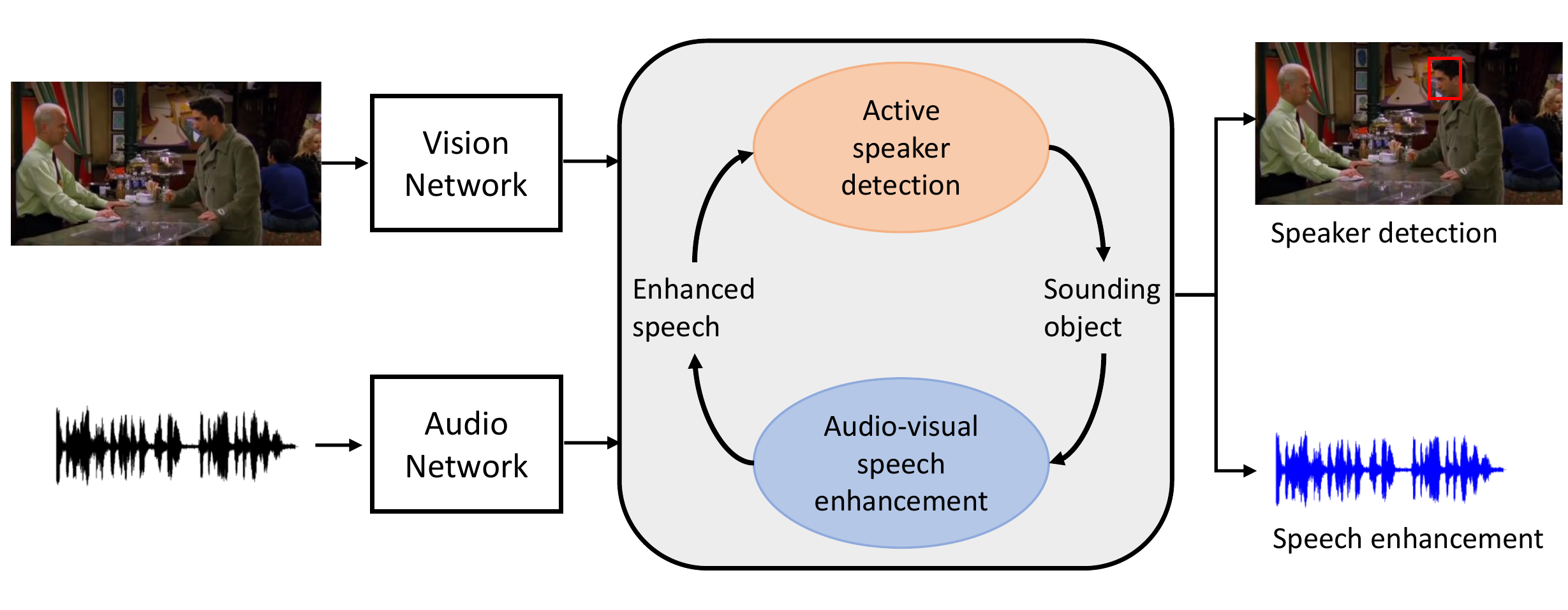}
	\caption{We propose a unified learning framework to jointly learn active speaker detection and audio-visual speech enhancement.}
	\label{fig:SmallModel}
\end{figure}

To further achieve the mutual learning of both audio enhancement and visual detection, a scheme of cross-modal circulant fusion is also proposed to leverage the complementary cues between the bifurcated processes for the establishment of their associations. 
The more accurate the detection result, the more reliable visual information to guide the speech enhancement; the cleaner the enhanced sound, the more distinctive the audio embedding, which will in turn help the detection, and overall performance can be guaranteed in such cyclic mutual learning as shown in Fig. \ref{fig:SmallModel}. The main contributions of this work can be summarized as:

(1)A novel unified framework ADENet is proposed to achieve mutually beneficial between audio enhancement and visual detection based on   audio-visual correlation learning.

(2)A cross-modal conformer is introduced to exploit diverse correlations between auditory and visual modalities, which is able to overcome the distribution misalignment of multi-modal features with a variant of layer normalization.



(3)A cross-modal circulant fusion scheme is also proposed to enable intrinsic assistance of two tasks by audio-visual feature interaction.

%

(4)Experiments on 3 benchmark datasets demonstrate a superior performance compared to  other state-of-the-art methods for active speaker detection and audio-visual speech enhancement in challenging scenarios.

\section{RELATED WORK}
\label{sec:related_work}

\subsection{Active Speaker Detection}
\label{ssec:subhead}

Active Speaker Detection(ASD) is to find who is speaking in a video clip that contains more than one speaker. In most realistic scenarios, the sound/voice might come from multiple speakers, which is challenging in finding a specific speaker. Earlier methods achieve this purpose by leveraging the audio cues of the voice activity detector to determine the presence of speech \cite{moattar2009simple,6737222, patrona2016visual}. With the development of convolutional neural networks, visual information has been taken full advantage of into the active speaker detection task,  $e$.$g$., a speaker can be found by analyzing the facial or lip movements \cite{haider2016active}. But for some non-speaking activities, $e$.$g$., eating food, chewing and grinning, active speaker detection has not fully benefited from in-depth modeling due to the lack of audio information. This also means that the visual or audio mode alone can not complete this task favourably.

Recent studies have shown a significant improvement by audio-visual modeling in different tasks, such as emotion recognition \cite{martin2006enterface, wu2013two, nie2020c}, speech recognition\cite{neti2000audio, tao2020end, liu2020re}, and etc. As the speech rhythm and word pronunciation are closely correlated with facial motion, the audio-visual based fusion scheme is a feasible and promising solution to the task of active speaker detection. Roth $et$ $al.$ \cite{roth2020ava} proposed the AVA-ActiveSpeaker dataset, the first large-scale video benchmark for the active speaker detection task. With this dataset, the author also introduced a baseline model based on a two-stream network including an audio-visual feature extract backbone and a prediction network in an end-to-end fashion. Since then, more and more works began to learn from this structure. Considering the temporal dependency between audio and visual, some works employed a temporal structure to build their prediction networks, such as recurrent neural network (RNN) \cite{tao2017bimodal,tao2019end}, gated recurrent unit (GRU) \cite{roth2020ava}, long short-term memory (LSTM) \cite{sharma2020crossmodal, shvets2019leveraging}, and transformer layer \cite{tao2021someone}.

Since these prediction networks separately encode the unimodal features of audio and video, the cross-modal synchronization information has not been fully exploited in audio-visual feature extraction. For visual features, recent works have been unified by using 3D CNN to extract the visual information and temporal dependency from video\cite{kopuklu2021design, tao2021someone}. For audio features, using Convolutional Neural Networks (CNNs) with log-Mel\cite{tao2021someone} or Short-Time Fourier Transform (STFT) spectrograms as inputs, can make a great success in traditional speech and audio processing. Based on the interpretation of CNNs as a data-driven filter-bank, another strategy is to apply CNNs directly on the audio waveforms to capture discriminative information\cite{kopuklu2021design}. Since the extracted features by these models contain plenty of noise interference, which is still a challenge to affect the performance of detection. As an inspiration, a speech enhancement structure is proposed to purify the background noise in the features to assist the detection task.

\subsection{Audio-Visual Speech Enhancement}
\label{ssec:subhead2}

Speech enhancement, a core mission of audio separation, is to extract a clean target speaker's utterance from a mixed audio signal. Generally, there are two ways to enhance the speech quality \cite{ito2021audio}. One is to utilize the speaker embedding for conditional modeling, the other is to make use of the visual modality. The studies of speech perception have shown that watching a speaker's facial movements could dramatically improve the capability of speech recognition for a target speaker in a noise environment\cite{golumbic2013visual}. Current deep audio-visual speech enhancement methods typically focus on exploiting the close connection between audio-visual streams\cite{gabbay2017visual, ephrat2018looking, afouras2018deep, morrone2019face, ito2021audio, afouras2018conversation,afouras2020self, owens2018audio, aldeneh2021role,gao2021visualvoice}, and these approaches can be categorized by the means how they process visual information. In specific, some methods combine the detection and tracking of faces to guide the enhancement process\cite{gabbay2017visual, afouras2018deep, morrone2019face, ephrat2018looking,gao2021visualvoice}, while the others discard the paradigm of detection and directly process raw video frames without any preprocessing \cite{afouras2020self, owens2018audio}. Both the ways have obtained satisfactory performance by learning the association between visual motions and speech content. To further achieve a considerable performance in realistic environments, these methods also utilize reliable visual information, such as lip movements, to avoid inaccurate extraction of human voices.

However, if the moving visual object is silent ($e$.$g$., mute chewing), these models usually fail to isolate all-zero sound for it. For misleading visual information, more recent studies incorporate a landmark-based facial embedding to perform speech enhancement\cite{Morrone2019}. When only in simple audio-visual scenarios, learning visual embedding from facial landmarks can benefit the process of enhancement. In another different way, we overcome the current deficiencies of AVSE by using the superiority of ASD, which can detect speaking objects to guide speech enhancement more effectively.






\subsection{Audio-Visual Learning}
\label{ssec:subhead3}

By exploiting the synchronized links \cite{KING2005R339, 8403294, li2016weakly, li2016multimedia} between visual and auditory modality, audio-visual learning strategy has been employed in a number of interesting tasks: audio-visual action recognition \cite{gao2020listen, kazakos2019epic}, emotion recognition \cite{martin2006enterface, wu2013two, nie2020c}, audio-visual event localization \cite{hu2021class, rao2021decompose}, and talking face generation \cite{zhou2021pose, wang2021one}. Unlike all these single-task approaches, our work addresses active speaker detection and audio-visual speech enhancement simultaneously, and proposes a novel unified framework for both detection and enhancement tasks, which exploits the interaction between audio and visual to make them mutually beneficial.

\section{Method}
\label{sec:method}
\begin{figure*}[tbp	]
	\centering
	\includegraphics[width=18cm]{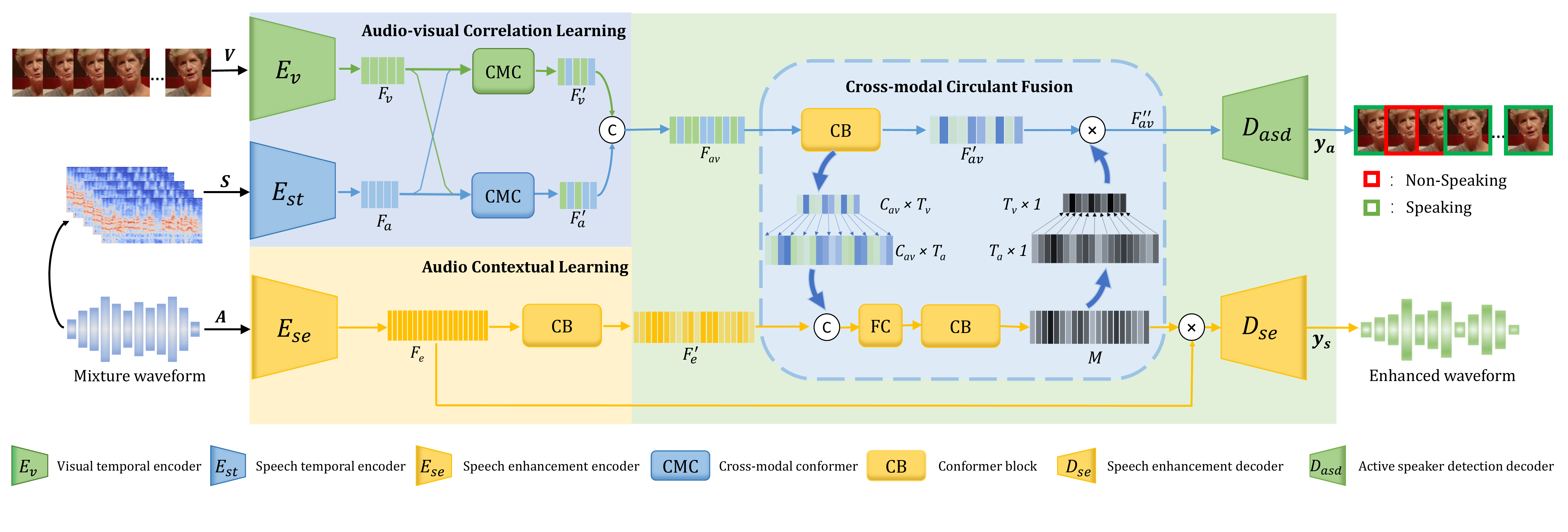}
	\caption{\textbf{The overall pipeline of our proposed ADENet model.} Its framework is divided into three stages: audio-visual correlation learning, audio contextual learning, and cross-modal circulant fusion. The audio-visual correlation learning and the audio contextual learning aim at modeling the associations between multi-modal data and extracting contextual embeddings in the audio domain, respectively. Then, cross-modal circulant fusion is proposed to integrate correlation features and contextual features for active speaker detection and speech enhancement.  }
	\label{fig:ADENet}
\end{figure*}

In this section, the proposed model ADENet is introduced elaboratively, which is an end-to-end unified framework. As shown in Fig.\ref{fig:ADENet}, the mixed audio signals, as well as the continuous face clip frames are taken as input for ADENet, and the outputs are composed of active speaker tags and enhanced target speech. The overall model is divided into three parts. We first explore an efficient feature extraction formulation by learning audio-visual correlation. Then, we provide contextual information from time-domain audio signals. To further make the two tasks mutually beneficial, the audio-visual correlation cues and audio contextual information are integrated by the cross-modal circulant fusion. Finally, we introduce our model predictions and training criteria.

\subsection{Audio-Visual Correlation Learning}
\label{sec:correlation}

Our audio-visual correlation learning is composed of an audio-visual encoding frontend and a relational modeling backend, as illustrated in Section \ref{sec:intro}. The frontend contains a speech temporal encoder and a visual temporal encoder. Their function is to encode the frame-based input audio and video signals into the time sequence of audio and video embeddings, which represent the temporal context. The backend consists of a cross-modal conformer to learn the correlations between audio and video features, and a multi-modal layer normalization to solve inconsistent distribution of features in different modalities.

\noindent\textbf{Audio-Visual Temporal Encoder.} Given an input video clip, the face tracks are firstly cropped for each candidate speaker and reshaped into a specified size, $e$.$g$., $112\times112$. Similarly, we extract the corresponding audio signal and represent it by a vector of Mel-frequency cepstral coefficients (MFCCs).

\textit{Speech Temporal Encoder}: To learn semantic features containing crucial vocal activities from the short-term audio content,
we introduce a 2D ResNet34 network with squeeze-and-excitation (SE) \cite{hu2018squeeze} module, which takes 13-dimensional MFCCs of the windows as input and outputs a 128-dimensional audio embedding $F_a$. The ResNet34 network can effectively strengthen the weight of key features in MFCCs by means of SE block, which would facilitate the subsequent audio-visual feature interactions. For example, the speech temporal encoder $E_{st}$ can pay more attention to audio clips with voice activity.

\textit{Visual Temporal Encoder}:  Different from the majority of state-of-the-art approaches \cite{alcazar2020active, roth2020ava, chung2019naver} that only apply 2D CNN as a visual encoder, we choose to combine 3D CNN and 2D CNN together. This idea is inspired by the advantage of 3D networks in capturing motion patterns of face crops, which is the essentially indicative information of active speakers.  

Therefore, we design a visual temporal encoder $E_v$, which is composed of the visual frontend and the visual temporal network. The former consists of a 3D convolutional layer (3D Conv) followed by a ResNet18 block \cite{afouras2018conversation}, which encodes the video frame stream into a sequence of frame-based embedding. The latter is a visual temporal network consisting  of a video temporal convolutional block (V-TCN), which has five residual connected rectified linear unit (ReLU), batch normalization (BN) and depth-wise separable convolutional layers (DS Conv1D) \cite{afouras2018deep}, followed by a Conv1D layer to reduce the feature dimension. It aims to extract the visual spatio-temporal information from the video segments. After the above processing, we can get the final face motion features $F_v$.

\noindent\textbf{Cross-Modal Conformer.}
Considering the information from different modalities can compensate each other in audio-visual tasks, we propose a cross-modal conformer based on the state-of-the-art ASR encoder architecture conformer block\cite{gulati2020conformer} to make better use of the correlations between audio and visual features. This architecture has demonstrated its effectiveness in temporal information processing\cite{zhangictcas}. The proposed cross-modal conformer is composed of five modules stacked, $i$.$e$, a feed-forward network module, a cross-modal attention module, a convolution module, a second feed-forward network module, and a layer normalization in the end.

\begin{figure}
	\centering
	\includegraphics[width=0.4\textwidth]{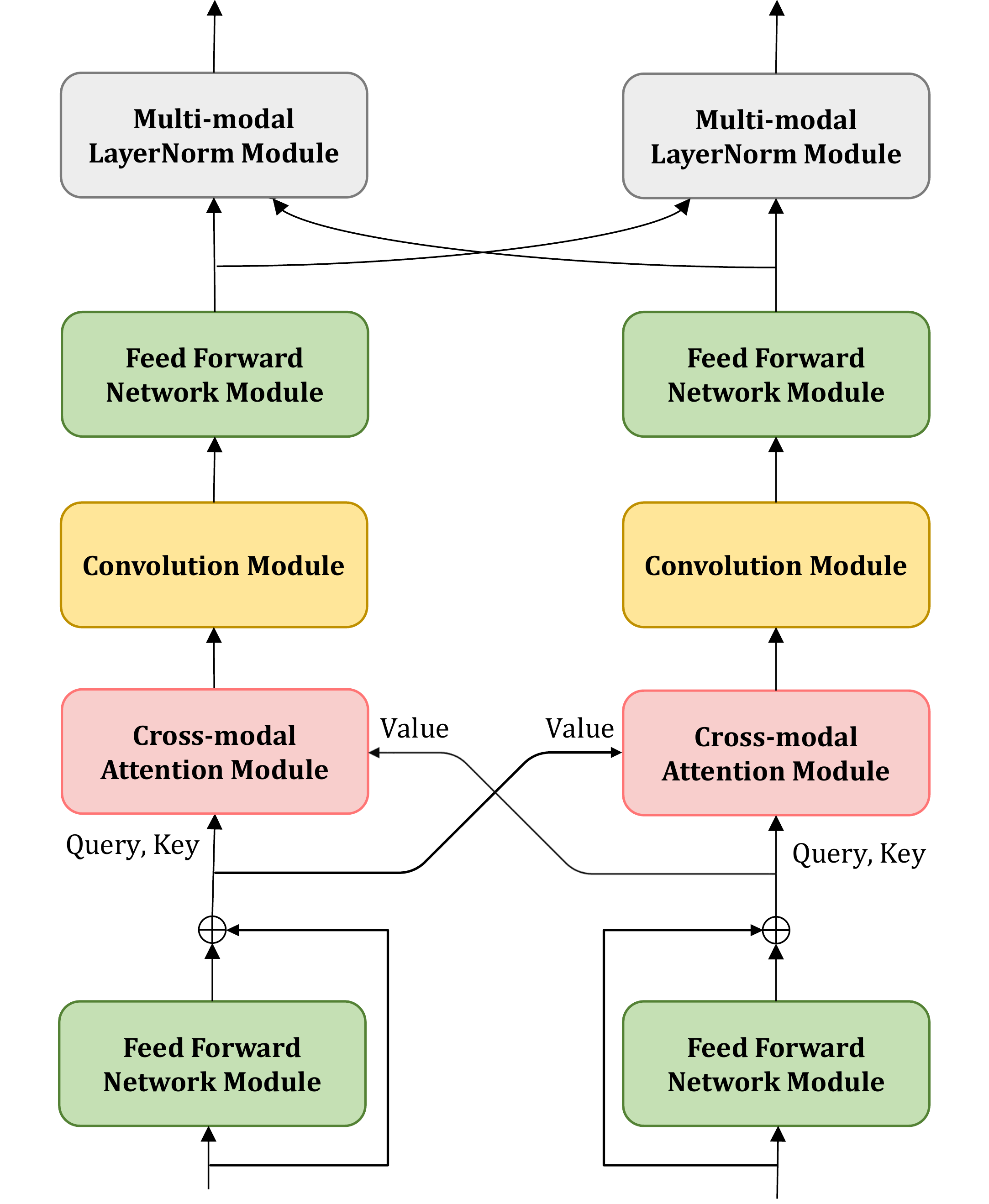}
	\caption{\textbf{Cross-modal conformer.} The conformer block modules consists of feed forward network(FFN), cross-modal attention(CMA), convolution module(Conv), and multi-modal layer normalization(MLN).}
	\label{fig:CMC}
\end{figure}

During processing, the first feed-forward network(FFN) takes the features of the audio-visual encoder as input, which represent the speaker's activities for audio $F_a$ and video $F_v$, respectively, and outputs the processed audio feature $\overline{F_a}$ and visual feature $\overline{F_v}$. Then, as shown in Fig.\ref{fig:CMC}, a cross-modal attention layer (CMA) takes $\overline{F_a}$ and $\overline{F_v}$ as input, which can be define as:

\begin{equation}
	\begin{aligned}
		& \overline{F_{a}'} = softmax(\frac{Q_a  K_a^{T} }{\sqrt{d}}  )V_v \\
		& \overline{F_{v}'} = softmax(\frac{Q_v  K_v^{T} }{\sqrt{d}}  )V_a \\
	\end{aligned}
\end{equation}
\begin{alignat}{2}
	Q_a = \overline{F_a} W_{A1} &\quad K_a = \overline{F_a} W_{A2} &\quad V_a = \overline{F_a} W_{A3} \\
	Q_v = \overline{F_v} W_{V1} &\quad K_v = \overline{F_v} W_{V2} &\quad V_v = \overline{F_v} W_{V3}
\end{alignat}
\noindent where $W_{A1}$, $W_{A2}$, $W_{A3}$, $W_{V1}$, $W_{V2}$ and $W_{V3}$ are the weights of the linear layers, $d$ is the feature dimension. We regard the attention $ \overline{F_{x}'}, x\in \{a,v\} $  as a probability distribution that illustrates how much an audio embedding responds to a candidate speaker in the temporal dimension, which means that a high probability indicates a high correlation between the audio and the speaker.

\begin{figure}[t]
	\hspace{-1cm}
	\centering
	\subfigure[\textbf{Unaligned}]{
		\begin{minipage}[t]{0.45\linewidth}
			\centering
			
			\includegraphics[width=1.1\linewidth]{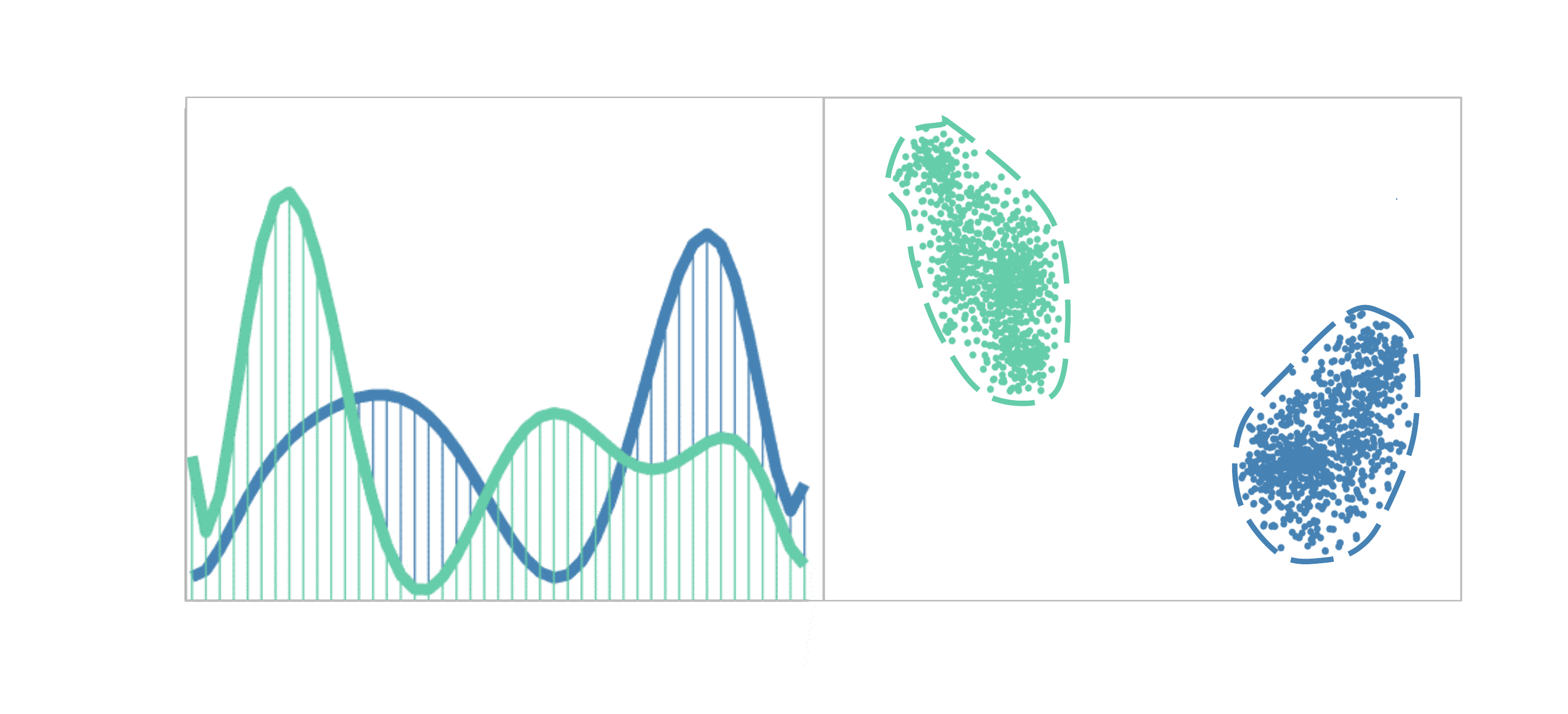}
		\end{minipage}
	}
	\subfigure[\textbf{Aligned}]{
		\begin{minipage}[t]{0.45\linewidth}
			\centering

			\includegraphics[width=1.1\linewidth]{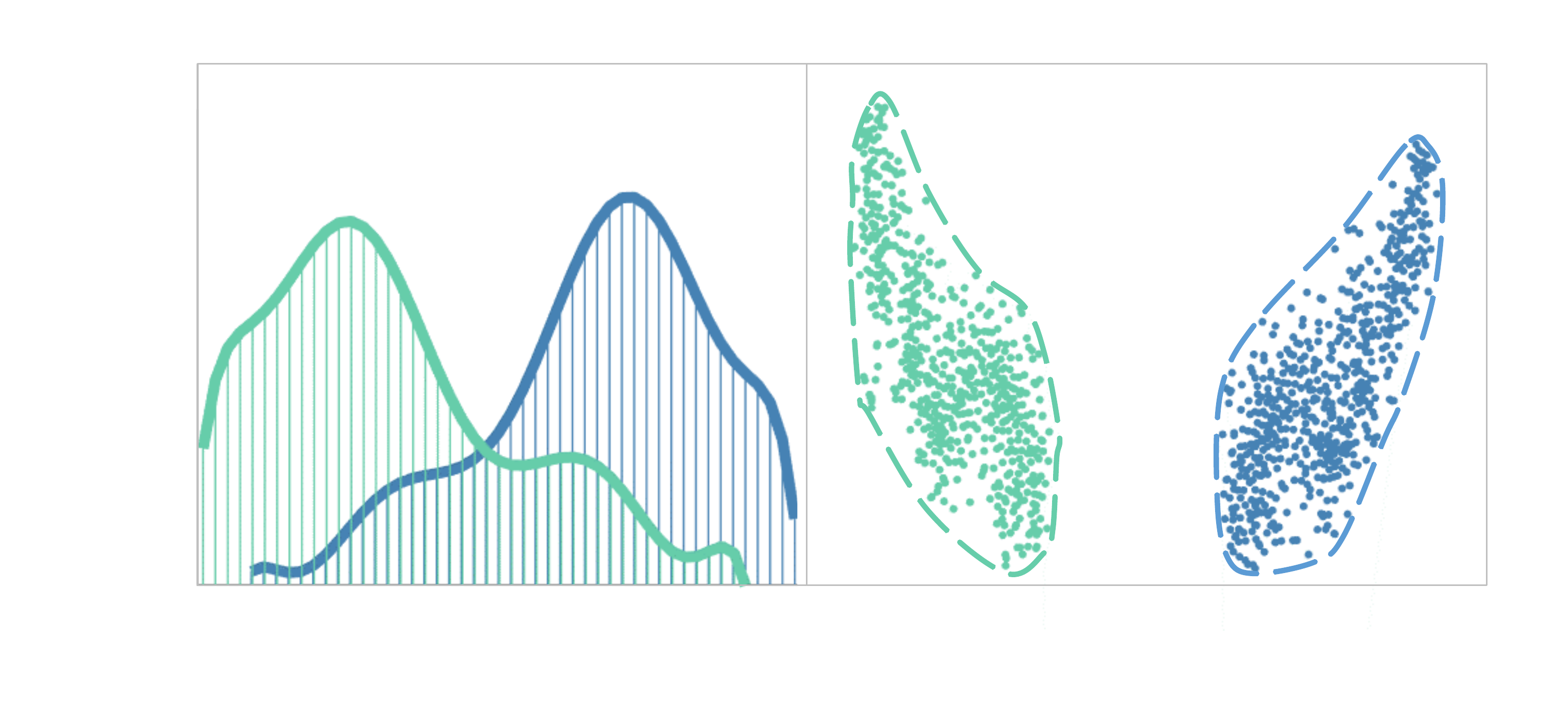}
		\end{minipage}
	}
	\caption{\textbf{Feature Distribution Alignment.} (a) shows the distribution and topology of \textcolor{Blue}{audio} and \textcolor{Green}{visual} features. It is obviously that visual features are not aligned with audio ones. Thanks to the multi-modal layer normalization, visual and auditory features are brought into similar distributions as (b). }
	\label{fig:MLN}
\end{figure}

After the operation of attention modulation, the features are processed in the sequence of a convolution module, a second feed-forward network module and a layer normalization. Although the cross-modal conformer has successfully established the similarity between audio and visual modalities, it is not able to solve the distribution misalignment among different domains (see Fig. \ref{fig:MLN} (a)). Therefore, we introduce a multi-modal layer normalization (MLN) to address this problem.
Fig. \ref{fig:MLN} shows our model's learned audio and visual features via 2D TSNE projections \cite{van2008visualizing}, and the distribution characteristics of these features via matplotlib library \footnote{https://matplotlib.org/stable/index.html}.

\noindent\textbf{Multi-Modal Layer Normalization.} Layer Normalization\cite{ba2016layer} has always been an indispensable part of the conformer. Given an input tensor $x\in \mathbb{R}^{N \times C \times T}$, LN can be formulated as:

\begin{equation}
	\begin{aligned}
		& u(x^{l}) =\frac{1}{T} \sum_{i=1}^{T}x_{i}^{l}  \quad
		\sigma (x^{l})=\sqrt{\frac{1}{T}\sum_{i=1}^{T}(x_{i}^{l}-u(x^{l}))^{2}} \\
		& LN(x^{l})= \gamma \left(\frac{x^{l}-u(x^{l})}{\sigma(x^{l})} \right) + \beta
	\end{aligned}
\end{equation}

\noindent where $N,C$ represent the number of batch-size and channel,  $T$ denotes the number of hidden units in a layer, $u(x^{l})$ and $\sigma (x^{l})$ are the mean and standard deviation for the $l^{th}$  layer of feature map $x$, and $\gamma, \beta \in \mathbb{R}^{C}$ are defined as the gain and bias parameters.

As a normalization approach widely used for single modality, LN is difficult to balance data relationships for multi modalities because the inherent distribution inconsistency would make data fail to be assembled into a pleasing representation. Also as shown in Fig. \ref{fig:MLN}(a), visual features and audio features are out of balance without any form of distribution alignment, which would finally lead to a performance degradation \cite{michelsanti2021overview}.

To solve this problem, a multi-modal layer normalization (MLN) is proposed as a simple and efficient variant to LN to align the distributions of two modalities. The MLN receives a tensor $x\in \mathbb{R}^{N \times C \times T}$ and a constraint variable $y\in \mathbb{R}^{N \times C \times T}$ as input, which can be defined as:

\begin{equation}
	\begin{aligned}
		& MLN(x^{l}, y)= \gamma \left(\frac{x^{l}-u(x^{l})}{\sigma(x^{l})}+tanh(f(y)) \right) + \beta \\
	\end{aligned}
\end{equation}

\noindent where $f(.)$ is a linear layer the represents affine transformation, and the tanh function is to constrain the range of units to [-1,1]. In the implementation of MLN, the tensor $x$ and the constraint variable $y$ can be audio features and visual features respectively, or vice versa. In this way, our MLN is able to ensure the distribution alignment among different modalities, as shown in Fig. \ref{fig:MLN}(b).

\subsection{Audio Contextual Learning}
\label{sec:audio_context}
Different from the speech temporal encoder designed in Section \ref{sec:correlation} that concentrates on crucial voice activities in short-term speech segments, the purpose of audio contextual learning network mainly focuses on extracting features with contextual information, which usually requires input data with long-term temporal dependencies. Considering the high temporal resolution of the raw audio signals \cite{oord2016wavenet}, we intend to use the raw audio waveforms as the input to the network.

As shown in Fig.\ref{fig:ADENet}, we follow the architecture of Conv-TasNet\cite{luo2019conv} for the design of the proposed module,  which consists of two processing stages: an encoder and a separation network.
The speech enhancement encoder is a 1-D convolution layer with a kernel size of $L$ and stride size of $L/2$, which can be characterized as a filter-bank. It takes the mixture waveform $\textbf{A} \in \mathbb{R}^{1\times T}$ as input, and transforms $\textbf{A}$ to a feature map $F_e \in \mathbb{R}^{C_{se}\times T_a}$, where $T$ is the length of mix audio, $C_{se}$ and $T_a$ are the channel and temporal length of the feature vectors respectively.

Inspired by the outstanding performance of conformer in speech processing\cite{gulati2020conformer}, the original temporal convolution network(TCN) in the separation network is replaced by a conformer block to allow direct context-aware modeling on the speech sequences. The speech separation network composed of a series of conformer blocks can efficiently capture both local and global context information, while estimating a contextual feature $F'_e \in \mathbb{R}^{C_{se}\times T_a}$ as output. Then, the embedding $F'_e$ is sent to the next cross-modal circulant fusion module for multi-task interaction.

\subsection{Multi-Task Interaction}
Based on the audio-visual correlation learning and audio contextual learning above, the aligned correlation embedding and the contextual embedding can be obtained. The former type of embedding has audio-visual utterance-level synchronization characteristics, which is suitable for active speaker detection, and the latter is beneficial to information transmission in speech enhancement via context-aware modeling.
Then to learn both active speaker detection and audio-visual speech enhancement, we design a cross-modal circulant fusion strategy in a pseudo cycle manner by mapping and combining two types of embeddings as an interactive compensation.

\noindent\textbf{Cross-Modal Circulant Fusion.} The process of cross-modal circulant fusion starts with $F_{av}$, which contains multi-modal temporal information from audio-visual correlation learning. Then, the utterance-level temporal information is modeled by a pure conformer block in audio-visual correlation representation, and the corresponding output $F'_{av}$ contains the motion status the speaker in the sync frame. In addition, to improve the influence of the active speaker frame on the contextual embedding, we firstly use linear up-sampling on the correlation feature to align with contextual embedding $F'_e$, and set the sampling scale to 32. The fusion process is then performed through a simple concatenation operation over the channel dimensions of two features, followed by a fully connected layer to reduce the feature dimension.
Finally, the aggregate features are fed through a conformer block to estimate the target masks $M$, which can be defined as:
\begin{equation}
	M = ReLU(CB(FC([F'_e; Up(F'_{av})])))
\end{equation}
where $CB$ and $FC$ denote the conformer block and fully connect layer, respectively, and $ReLU$ is to ensure that the mask is non-negative. The target mask $M$ represents the weight of the clean speech, which is important for the task of speaker detection to avoid the interference of noise.

To make the results of ASD more accurate, we combine $F'_{av}$ with the max pooled mask via a dot product operation, as formulated below:

\begin{equation}
	F''_{av} = MaxPool(M)  \otimes F'_{av}
\end{equation}

With the help of cross-modal circulant fusion, the speech mask of the target speaker can be estimated and more effective correlation embeddings can be learned collaboratively.

\subsection{Decoder}

\noindent\textbf{Active Speaker Detection Decoder.} To distinguish the speaking and non-speaking video frames, we build a predictive decoder $D_{asd}$, which contains a linear layer, to project the joint audio-visual feature to the prediction sequence. The output  of the decoder is passed via sigmoid activation to produce the expected label values $\textbf{y}_a$ (between 0 and 1), which indicate each candidate's probability of being the active speaker.
\begin{equation}
	\textbf{y}_a = D_{asd} (F''_{av})
\end{equation}

\noindent\textbf{Speech Enhancement Decoder.} To obtain the masked audio representation, the learned speech mask of the target speaker $M$ is applied to audio contextual embedding $F_e$ by the operation of dot product. In our speech enhancement decoder, a 1-D transposed convolution module is used to reconstruct the enhanced speech signals for the masked audio feature:

\begin{equation}
	\textbf{y}_s = D_{se} (M \otimes F_e)
\end{equation}
\noindent where $\textbf{y}_s$ is enhanced waveform, $D_{se}$ is speech enhancement decoder, and $\otimes$ denotes dot product operation.

\subsection{Multi-Task Learning Objectives}

Our training strategy is described into two different criteria for speech enhancement $L_{se}$ and active speaker detection $L_{asd}$. 

\noindent\textbf{Speech Enhancement Loss:} As formulated in Eq.\ref{eq_se}, the training criterion for the speech enhancement is based on a scale-invariant signal-to-noise radio(SI-SDR) between the extracted and clean target speech,  $\hat{\textbf{y}_s}$ is the ground truth target speech, $\textbf{y}_s$ is the extract speech, and $||\textbf{y}_s|{{|}^{2}} = \langle {\textbf{y}_s}, {\textbf{y}_s} \rangle $ denotes the signal power. Scale invariance is ensured by normalizing $\hat{\textbf{y}_s}$ and $\textbf{y}_s$ to zero-mean prior to the calculation. 
\begin{equation}
	\label{eq_se}
	\left\{
	\begin{array}{l}
		\textbf{s}_{target} = \dfrac{\langle\hat{\textbf{y}_s},\textbf{y}_s\rangle\textbf{y}_s}{||\textbf{y}_s|{{|}^{2}}},   \\
		\\
		\textbf{e}_{noise} = \hat{\textbf{y}_s} - \textbf{s}_{target},    \\
		\\
		{{L}_{se}}=-20{{\log }_{10}}\dfrac{||\textbf{s}_{target}||}{||\textbf{e}_{noise}|| }
		
	\end{array}
	\right.
\end{equation}

\noindent\textbf{Active Speaker Detection Loss:} To accurately predict who is speaking in a video clip, we regard active speaker detection as a frame-level classification task.The predicted label sequence is computed with the ground truth label sequence by cross-entropy loss. The loss function is presented in Eq.\ref{eq_asd}, 

\begin{equation}
	\begin{array}{l}
		\label{eq_asd}
		{{L}_{asd}}=-\frac{1}{T_v}\sum\limits_{i=1}^{T_v}{({\hat{\textbf{y}_{a}}^{i}}\cdot \log {\textbf{y}_{a}^{i}}+(1-{\hat{\textbf{y}_{a}}^{i}})\cdot \log (1-{\textbf{y}_{a}^{i}}))} \\
	\end{array}
\end{equation}
\noindent where $\textbf{y}_{a}^{i}$ and $\hat{\textbf{y}_{a}}^{i}$ are the predicted and the ground truth ASD labels of $i^{th}$ video frame, $i \in [1,T_v]$.  $T_v$ refers to the number of video frame.

The overall objective function for training $L$ can be defined as follow:
\begin{equation}
	\label{eq_total_loss}
	L = \lambda_1 L_{se} + \lambda_2 L_{asd}
\end{equation}
\noindent where $\lambda_1$ and $\lambda_2$ are the parameters controlling their relative importance, respectively.

\section{EXPERIMENTS}
\subsection{Datasets}
\begin{figure}
	\hspace{-0.5cm}
	\centering
	\includegraphics[scale=0.9]{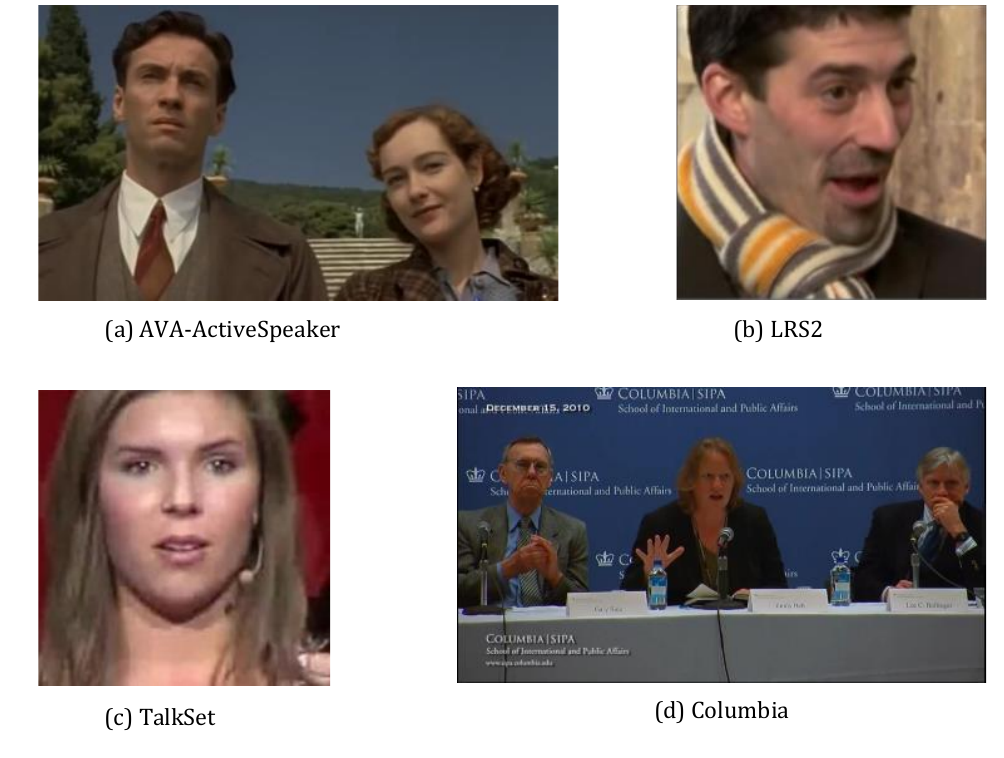}
	\vspace{-20pt}
	\caption{Example frames from the datasets used in this paper.}
	\label{fig:data}
\end{figure}

\noindent\textbf{AVA-ActiveSpeaker dataset} \cite{roth2020ava} (Fig. \ref{fig:data}a) is the first large-scale standard benchmark for active speaker detection in the wild. It contains 262 YouTube videos from film industries around the world, 120 of which are used for training, 33 for validation, and 109 for testing. Each video is annotated from minutes 15 to 30, recorded at 25-30 fps, and cropped into 1 to 10 seconds video utterances.

\noindent\textbf{Lip Reading Sentences 2 (LRS2)} \cite{afouras2018deep} (Fig. \ref{fig:data}b) is a widely-used benchmark for audio-visual speech enhancement. It contains 224 hours of in-the-wild video clips, which are all from BBC television programs. Each video clip often contains only one speaker.

\noindent\textbf{TalkSet} \cite{tao2021someone} (Fig. \ref{fig:data}c) is a mixture of two datasets, Voxceleb2 \cite{chung2018voxceleb2} and LRS3 \cite{afouras2018lrs3}, for the active speaker detection task. The dataset contains 90,000 videos with active voice from VoxCeleb2 and 60,000 videos without active voice from LRS3.

\noindent\textbf{Columbia} \cite{chakravarty2016cross} (Fig. \ref{fig:data}d) is another standard benchmark test dataset for active speaker detection. It contains an 86-minute panel discussion, with multiple individuals taking turns in speaking, in which 2-3 speakers are visible at any given time.

\noindent\textbf{MUSAN corpus} \cite{snyder2015musan} is a noise dataset, which consists of over 900 noises, 42 hours of music from various genres and 60 hours of speech from twelve languages. It is usually used for speech enhancement tasks as a real noise background, or mixing with training data to the strength of regularization and improve the generalization performance.

\subsection{Data Preprocess.}

\noindent\textbf{Audio Data Preprocess.} As in \cite{tao2021someone}, randomly mix a sample of the batch  with another negative sampling for data argumentation. By extracting 13-channel filterbanks features computed from a 25ms window with a stride of 10 ms, the obtained MFCCs are taken as the input of the speech temporal encoder. Another speech enhancement encoder takes the mixed audio signal as input.

\noindent\textbf{Visual Data Preprocess.} All training videos are firstly segmented into shots with FFmpeg  \cite{lienhart2001reliable}, and the output images are cropped out the faces with a resolution of 112$\times$112. The obtained faces are then randomly rotated and flipped to perform visual augmentation. During the testing phase, the image size is only reshaped to 112$\times$112 without any augmentation.

\subsection{Train/test split and Evaluation}
For large-scale AVA-ActiveSpeaker dataset, the official split (train/validation/test) is adopted. The test set is unavailable (contest exclusive) and held out for the ActivityNet challenge, which makes us conduct the experiments on the validation set, as the previous works \cite{alcazar2020active, chung2019naver}. For other datasets, we directly use the same split setting as provided by the official.

To compare with the prior methods, five common-used metrics are employed for active speaker detection and speech enhancement evaluation. For detection, we choose to mean precision (mAP), area under the ROC curve (AUC) and F1 to measure the model performance. For speech enhancement, the metrics of SDR and perceptual evaluation of speech quality (PESQ) \cite{rix2001perceptual} are employed to reflect the quality of enhanced speech signals.

\subsection{Implementation Details}
\noindent\textbf{Audio-Visual Correlation Learning.} For the speech temporal encoder, we use four squeeze-and-excitation convolution layers with kernel sizes 3, 4, 6 and 3. The number of filter channels is set to 16, 32, 64 and 128, respectively.
To synchronize speech features, the output dimension of the visual temporal encoder is set to 128. Besides, the attention head is set to 8 in the cross-modal conformer.

\noindent\textbf{Audio Contextual Learning.} In the speech enhancement encoder, we set the kernel size K to 40 and stride S to 20. The number of attention heads in the conformer block is set to 8 as well.

\noindent\textbf{Cross-modal Circulant Fusion.} We set both the up-sampling and down-sampling multiples to 32.

\noindent\textbf{Training Details.} Our unified framework is implemented using PyTorch library \footnote{https://github.com/pytorch/pytorch}. The entire ADENet model is trained using an Adam optimizer \cite{kingma2014adam} with weight decay of 0.0001. The initial learning rate is $10^{-4}$, and we decrease it by 5$\%$ for every epoch. For training loss, $\lambda_1$ and  $\lambda_2$ are both set to 1.0 in Eq.\ref{eq_total_loss}. The computation platform is configured by a NVIDIA GeForce RTX 3090 GPU with 24 GB memory for all experiments.

\section{Result Analysis}
The evaluation of the proposed ADENet is performed from three points of view. (1) Performance comparison between the proposed baseline named ACLNet and the state-of-the-art is conducted for active speaker detection, while ablation studies are performed on the components of the ACLNet.
This baseline consists of an audio-visual correlation learning module, one conformer block and an ASD prediction decoder; (2) For the audio-visual speech enhancement task, we compare our model with the state-of-the-art methods on the LRS2 dataset, and qualitatively analyze the performance of our model; (3) Union-tasks testing is performed including detection and enhancement to verify the effectiveness of the entire framework.

\subsection{Active Speaker Detection}
\subsubsection{Comparison with the State-of-the-Art Methods}
Quantitatively evaluation for the active speaker detection based on audio-visual correlation learning, the proposed ACLNet is compared with the state-of-the-art methods on the AVA-ActiveSpeaker dataset.
For a fair comparison, we use the same evaluation protocol and the same metrics for these methods.
As shown in Table \ref{table-ASD-all}, the proposed model achieves 93.2$\%$ mAP and 97.2 $\%$ AUC, which favorably outperform the state-of-the-art methods without any pre-training, $i$.$e$., TalkNet\cite{tao2021someone}, by 0.9$\%$ mAP  and 0.4$\%$ AUC on the validation set. Meanwhile, we note that some methods \cite{leon2021maas, alcazar2020active, chung2019naver} utilize the pre-trained weights on other large-scale datasets. In contrast, our model only uses the AVA-ActiveSpeaker trainset to train the entire network from scratch without any additional processing.
\begin{table}[!htbp]
	\caption{Comparison with SOTA works on the AVA-ActiveSpeaker validation set.}
	\label{table-ASD-all}
	\begin{center}
		\resizebox{\linewidth}{!}{
			\begin{tabular}{l ccc }
				\toprule
				\textbf{Model} & \textbf{Pre-training?} & \textbf{mAP($\%$)} & \textbf{AUC($\%$)}  \\
				\midrule
				VGG-LSTM\cite{chung2019naver}& \Checkmark &85.1 & -  \\
				VGG-TempConv\cite{chung2019naver}& \Checkmark &85.5 & - \\
				ASC\cite{alcazar2020active} & \Checkmark & 87.1 & 86.76 \\
				MAAS-TAN\cite{leon2021maas} & \Checkmark &88.8 &  -\\
				UniCon\cite{zhang2021unicon} &  \XSolid & 92.2 & 97.0 \\
				TalkNet\cite{tao2021someone} & \XSolid & 92.3& 96.8 \\
				\textbf{ACLNet} & \XSolid & \textbf{93.2} & \textbf{97.2} \\
				\bottomrule
				
		\end{tabular}}
	\end{center}
\end{table}

By noticing that the model trained on the AVA-ActiveSpeaker dataset can not adapt to the real-world scene very well \cite{tao2021someone}, the proposed ACLNet model trained on TalkSet is then evaluated on the Columbia dataset to test its capability of cross-dataset generalization. The results are summarized in Table \ref{table-columbia}, where an experiment for 5 speakers in the scenario is presented. It reports the F1 score as the standard metric in this benchmark. For multiple-speaker videos, our approach also outperforms all previously reported results with a large margin. By evaluating both benchmark datasets, the overall performances also support the effectiveness and superiority of our audio-visual correlation learning strategy.
\subsubsection{Ablation Studies}
Experimental evaluations are performed to analyze the contributions of each modulized functionality in the proposed framework.

\noindent\textbf{The importance of audio-visual correlation information.} Ablation studies for audio-visual correlation learning are conducted to demonstrate the contribution of audio-visual relationship representations. To show the comparison results more intuitively, the performance is only evaluated on the ASD task. The experimental results are summarized in Table \ref{table-CMC}, which shows that, without multi-modal layer normalization, the detection performance drops 0.7$\%$ mAP and 0.4$\%$ AUC respectively on the AVA-ActiveSpeaker validation set. When the cross-modal conformer block is removed, the performance decreases to 92.1$\%$ mAP and 96.5$\%$ AUC. The results strongly prove the effectiveness of the cross-modal conformer and multi-modal layer normalization in learning the relationships between audio and visual modalities, as well as solving the distribution misalignment among them.

\begin{table}[!htbp]
	\begin{center}
		\caption{Ablation study of the cross-modal conformer and multi-modal layer normalization in ACLNet on the AVA-ActiveSpeaker validation set.}
		\label{table-CMC}
		\resizebox{\linewidth}{!}{
			\begin{tabular}{l cc }
				\toprule
				\textbf{Method} & \textbf{mAP($\%$)} & \textbf{AUC($\%$)}  \\
				\midrule
				w/o Cross-modal conformer & 92.1 & 96.5 \\
				w/o Mutil-modal layer normalization & 92.5& 96.8 \\
				\textbf{ACLNet} & \textbf{93.2} & \textbf{97.2} \\
				\bottomrule
		\end{tabular}}
	\end{center}
	
\end{table}

\noindent\textbf{The effect of different types of audio input for the audio-visual correlation learning.}
Selection  of different  types of audio input  has a significant impact on the performance of the model. 
As we discuss the multiple audio inputs  in Section \ref{sec:method}, we select the original audio signals and MFCCs features as the input of the audio-visual correlation learning, respectively. 
Since the inconsistent characteristics of different types of audio, we choose suitable audio feature extraction networks for each input.
Following recent works \cite{tao2021someone, 9580127}, we start from a baseline ACLNet using ResNet34-SE \cite{hu2018squeeze} as the speech temporal encoder, which takes MFCCs as audio input to the model. This baseline achieves 93.2$\%$ mAP and 97.2$\%$ AUC, respectively, as show in Table \ref{table-audio_type}. For the original audio signals, we replace the baseline's speech temporal encoder with the speech enhancement encoder mentioned in the Section \ref{sec:audio_context}, while keeping other modules the same. 
This resulted in the ACLNet's performance drops of 1.6$\%$ mAP and 0.8$\%$ AUC respectively. 
These experiments can effectively prove that using MFCCs features as audio input is beneficial to the audio-visual correlation learning.

\begin{table}[!htbp]
	\centering
	\caption{Ablation study of audio input in ACLNet on the AVA-ActiveSpeaker validation set.}
	\label{table-audio_type}
	
	\resizebox{0.85\linewidth}{!}{
		\begin{tabular}{cccc}
			\toprule
			\textbf{Audio Input} & \textbf{mAP($\%$)}  & \textbf{AUC($\%$)}  &  \\
			\midrule
			
			original audio signals         & 91.6 & 96.4 &  \\
			\textbf{MFCCs}             & \textbf{93.2} & \textbf{97.2} &  \\
			\bottomrule
	\end{tabular}}
\end{table}

\begin{figure*}[!h]
	\includegraphics[scale=0.62]{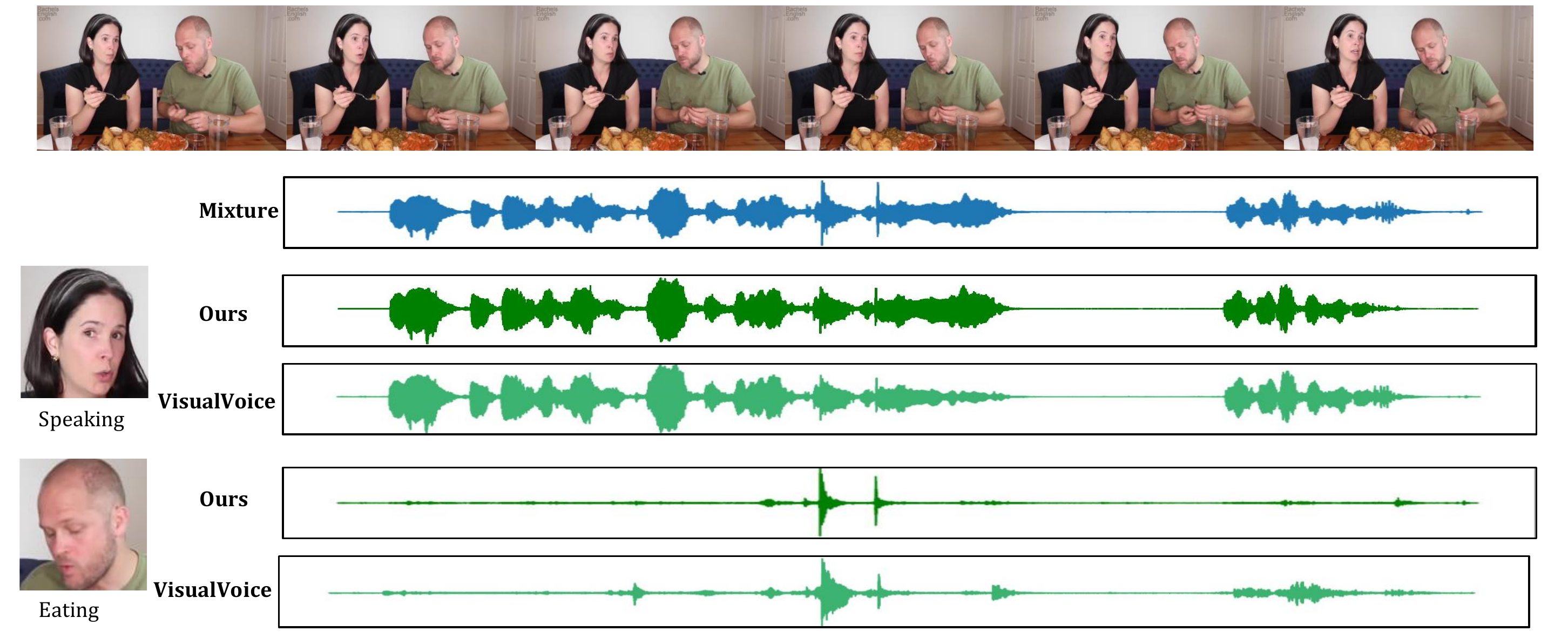}
	\vspace{-10pt}
	\caption{Qualitative results of audio-visual sound enhancement for both audible and silent objects.  The lady is speaking and the eating man is visible but not speaking. Our approach can simultaneously enhance individual sounds for for different objects, while the VisualVoice using the same training setting obtains poor enhancement results. }
	\label{fig:se_result}
\end{figure*}

\noindent\textbf{Different positions of multi-modal layer normalization in cross-modal conformer.} To further exploit where the multi-modal layer normalization may help most, we apply it to different positions in the cross-modal conformer. Table \ref{table-LN} shows that the proposed MLN is able to achieve the best performance improvement when it is in the position of LN. With this configuration, our model increases the overall mAP metric by 0.65\% and the AUC metric by 0.36\%. Besides, it is found that multi-modal layer normalization has different effects at different locations, which means that if the MLN is placed in another position of the module, it will reduce the performance of the model more or less. These results show that the MLN contributes the maximum functionality by replacing  LN in the last layer of conformers. Meanwhile, it re-emphasizes the advantages of MLN in the aligning distribution of audio-visual correlation features automatically and robustly.

\begin{table}[!tp]
	\begin{center}
		\caption{Performance evaluation of inserting multi-modal layer normalization into various positions of the cross-modal conformer. The top row is the performance of ACLNet without multi-modal layer normalization, which is used as the \underline{baseline result}.}
		\label{table-LN}
		\resizebox{\linewidth}{!}{
			\begin{tabular}{lccccccc }
				\toprule
				& \textbf{1st FFN} & \textbf{CMA} & \textbf{CONV}  & \textbf{2rd FFN} & \textbf{LN} & \textbf{mAP($\%$)} & \textbf{AUC($\%$)}  \\
				\midrule
				& & & & & & \underline{92.5}& \underline{96.8} \\
				\midrule
				& $\checkmark$  &&&& &  -0.02 & +0.0 \\
				& & $\checkmark$  &&& & -0.52 & -0.18 \\
				& & & $\checkmark$ && & -0.46 & -0.13 \\
				& & & & $\checkmark$ & & -0.04 & +0.0 \\
				& & & & & $\checkmark$ & \textbf{+0.65} & \textbf{+0.36} \\
				\bottomrule
		\end{tabular}}
	\end{center}
\end{table}

\begin{table}[!htp]
	\centering
	\caption{We show the audio-visual speech enhancement performance separately for testing examples where the objects is speaking (left) or No-Speaking (right).}
	\label{table-se}
	\resizebox{\linewidth}{!}{
		\begin{tabular}{c|cc|cc}
			\toprule
			\multicolumn{1}{c|}{ \multirow{2}{*}{\textbf{Method}} } & \multicolumn{2}{c|}{\textbf{Speaking Objects}} & \multicolumn{2}{c}{\textbf{No-Speaking Objects}}\\
			\cline{2-5}
			
			&\textbf{SDR(dB) $\uparrow$ } &\textbf{PESQ $\uparrow$} & \textbf{SDR(dB) $\uparrow$}&\textbf{PESQ $\uparrow$}\\
			\midrule
			VisualVoice \cite{gao2021visualvoice} & 10.1 & 2.8 & 0.3 & 1.7 \\
			\textbf{Ours} & \bfseries{10.5}& \bfseries{2.8} &\bfseries{4.8}&  \bfseries{1.9} \\
			\bottomrule
	\end{tabular}}
\end{table}

\begin{table*}[htb] 
	\begin{minipage}[]{0.55\textwidth} 
		
		\flushleft
		\caption{Comparsion with SOTA apporaches on the Columbia dataset for active speaker detection. F1 Scores (\%) for each speaker, and the overall average. }
		\scalebox{1.18}{
			\begin{tabular}{l ccccc|c }
				\toprule
				\multirow{2}{*}{\textbf{Method}} & \multicolumn{6}{c}{\textbf{Speaker}} \\
				& \textbf{Bell}  & \textbf{Boll} & \textbf{Lieb} & \textbf{Long} & \textbf{Sick} & \textbf{Avg}  \\
				\midrule
				Zach et al. \cite{zach2007duality} & 89.2 & 88.8 & 85.8 & 81.4 & 86.0 & 86.2 \\
				SyncNet \cite{chung2016out} & 93.7 & 83.4 & 86.8 & 97.7 & 86.1 & 89.5 \\
				LWTNet \cite{afouras2020self} & 92.6 & 82.4 & 88.7 & 94.4 & 95.9 & 90.8 \\
				S-VVAD \cite{shahid2021s} & 92.4 & \textbf{97.2} & 92.3 & 95.5 & 92.5 & 94.0 \\
				Truong et al. \cite{truong2021right} & 95.8 & 88.5 & 91.6 & 96.4 &  97.2 & 94.9 \\ 
				\midrule
				\textbf{ACLNet} & \textbf{97.4} & 88.1 & \textbf{97.5} & \textbf{98.5} & \textbf{98.0} & \textbf{95.9} \\
				\bottomrule
				
		\end{tabular}}
		\label{table-columbia}
		\label{table:by:fig} 
	\end{minipage}%
	\hspace{3mm}  %
	\begin{minipage}[]{0.4\textwidth} 
		\flushleft
		\caption{Comparsion with SOTA apporaches on LRS2 for speech enhancement.}
		\scalebox{1.15}{
			\begin{tabular}{l c|c}
				\toprule
				\textbf{Method} & \textbf{SDR(dB) $\uparrow$} & \textbf{PESQ $\uparrow$}   \\
				\midrule
				Deep-Clustering\cite{hershey2016deep} & 6.0 & 2.3 \\
				Conv-TasNet\cite{luo2019conv} & 10.7  & -  \\
				LWTNet \cite{afouras2020self} & 10.8 & 3.0  \\
				Afouras et al.\cite{afouras2018conversation} & 11.3 &  3.0 \\
				Truong et al. \cite{truong2021right} & 11.6 & 3.1 \\ 
				VisualVoice \cite{gao2021visualvoice} & 11.8 & 3.0 \\
				\midrule
				\textbf{Ours (w TCN)} & 12.3 & 3.0  \\
				\textbf{Ours (w Conformer)} & \textbf{12.9} & \textbf{3.1}  \\
				\bottomrule
				
		\end{tabular}}
		
		\label{table-lr2}
		\label{table:by:fig} 
	\end{minipage} 
\end{table*}

\begin{table*}[!ht]
	\begin{center}
		\caption{Comparsion with SOTA apporaches on the task of active speaker detection and speech enhancement in different signal-to-noise ratio environments.}
		\label{table:union-tasks}
		\vspace{-10pt}
		\resizebox{\linewidth}{!}{
			\begin{tabular}{l cccccccccccc}
				\toprule
				\multirow{2}*{\textbf{Method}} & \multicolumn{4}{c}{\textbf{0dB}} & \multicolumn{4}{c}{\textbf{5dB}} & \multicolumn{4}{c}{\textbf{10dB}}\\
				\cline{2-13}
				& \textbf{mAP} & \textbf{AUC} & \textbf{SDR} & \textbf{PESQ}  & \textbf{mAP} & \textbf{AUC} & \textbf{SDR} & \textbf{PESQ}  & \textbf{mAP} & \textbf{AUC} & \textbf{SDR} & \textbf{PESQ} \\
				\midrule
				
				TalkNet\cite{tao2021someone}& 89.1 &95.0 &- & - & 89.7 & 95.5 &- &- &91.1& 96.0 &-&-\\
				Conv-TasNet\cite{luo2019conv} & - & - & 8.7 & 3.0 & - & - & 11.1 & 3.2  & - & - & 13.8 & 3.5 \\
				VisualVoice\cite{gao2021visualvoice} &-&-&9.2&3.1&-&-&12.0&3.3& - & - & 14.3 & 3.5  \\
				TalkNet + VisualVoice & 89.1 & 95.0 & 10.4 & 3.2 & 89.7 & 95.5 & \textbf{13.2} & 3.4 & 91.1 & 96.0 & 14.8 & 3.6  \\
				\midrule
				\textbf{Ours (w/o A$\to$S)} &89.1 & 95.2 &10.4 &3.1 &90.2 & 95.6 &12.4 &3.3 &90.7 & 95.0 &15.1 &3.6  \\
				\textbf{Ours (w/o S$\to$A)} &89.0&95.0&10.3&3.1&88.8 &94.8 &12.7 &3.3 &91.1&95.5&14.9 &3.5  \\
				\textbf{Ours} & \textbf{89.7} & \textbf{95.9} & \textbf{10.6} & \textbf{3.2} & \textbf{90.3} & \textbf{96.2} & 12.8 & \textbf{3.4} & \textbf{92.1} & \textbf{96.9} & \textbf{15.2} & \textbf{3.7} \\
				\bottomrule
		\end{tabular}}
	\end{center}
\end{table*}

\subsection{Audio-Visual Speech Enhancement.}
\subsubsection{Comparison to State-of-the Art Methods} 
To quantitatively evaluate the audio-visual speech enhancement with ADENet model, we follow the protocol of \cite{afouras2020self} to create synthetic videos from LRS2 by combining a primary voice with another speaker's voice. Meanwhile, the SDR and PESQ are used as the validation metrics. However, the original LRS2 dataset does not provide ground truth labels of voice activity, which would make it difficult for the ADENet to achieve optimal performance. In order to provide accurate detection information for the audio-visual correlation learning module, we train this module with the pre-trained weights on the AVA-ActiveSpeaker dataset. And all the other modules in the ADENet are trained from scratch. Table \ref{table-lr2} has shown a large margin of performance between the proposed work and other methods.

Ablation studies are performed on the audio contextual learning network, which validate the effect of the separation network by adopting two functional models, $i$.$e$., TCN and conformer. It can be found that even though the incorporated TCN has obtained expected results by focusing on contextual information, the conformer block still can better perform context-aware modeling and further achieve overall performance enhancement. With both models, the proposed work outperforms VisualVoice \cite{gao2021visualvoice} and LWTNet \cite{afouras2020self} in SDR and PESQ.

To verify the ability of our approach to handle different scenarios, we conduct quantitative experiments on test videos based on whether the object is speaking or no-speaking in Table \ref{table-se}. For the speaking case, we use data of speaking targets with reliable visual cues in the video; for the no-speaking case, we pick out videos that contain the target lip movement without speaking, $e$.$g$., the mouth is chewing food. These silent data represent examples of misleading behavior in real-world scenarios. Table \ref{table-se} shows that in both scenarios, our method achieves the best enhancement results. Compared with VisualVoice, our approach is capable of identifying whether the candidate speaks, and employing speech enhancement for each speaker. The experimental results can well validate the superiority of the proposed  detection-enhancement strategy.

\subsubsection{Qualitative Results}
We further show qualitative enhancement results for speaking and silent objects in Fig. \ref{fig:se_result}. We can see that both VisualVoice and our method have similar performance for the speaking object, but our method has better enhancement results in handling the eating case. The results can illustrate that our model is more robust to  filtering noise from silent objects and can effectively perform audio-visual speech enhancement.

\subsection{Union-Tasks}
\subsubsection{Overall Performance}
Previous experiments have convincingly demonstrated the effectiveness of audio-visual correlation learning and audio contextual learning. Since the proposed unified framework aims to jointly learn active speaker detection and audio-visual speech enhancement, therefore, how to solve both tasks simultaneously and how they mutually benefit each other are addressed as well. 

For the evaluation of the two tasks, the performance comparison is performed with the other representative works such as TalkNet\cite{tao2021someone}, Conv-TasNet\cite{luo2019conv} and VisualVoice\cite{gao2021visualvoice}, which also have been trained and tested on the same dataset. 

In addition to the introduced these single-task models, a two-stage multi-task framework is also designed to jointly take an advantage of TalkNet and VisualVoice models. We firstly use TalkNet to determine whether the candidate is a speaker, if true, then perform the VisualVoice on the speaker for speech enhancement. If not, the enhancement is not needed.

To simulate the realistic scenarios containing noise, we employ the AVA-ActiveSpeaker dataset with MUSAN dataset as the background noise. Each method has been trained at three different signal-to-noise ratio configurations and evaluated using standard metrics, $i$.$e$., mAP, AUC, SDR, and PESQ. The results are summarized in Table \ref{table:union-tasks}. 
Based on the trained active speaker detection and audio-visual speech enhancement, the performance of our work outperforms the results of each single-task model by a large margin.  And we noticed that compared to the two-stage framework, our one-stage unified model has better overall performance. This also demonstrates that the idea of joint training for these two tasks is beneficial to establish more reliable cross-task associations.

Furthermore, to verify the positive effect of cross-modal circulant fusion strategy in both tasks, we set up two different ablation experiments, one without stream from ASD to SE (A$\to$S) and the other without stream from SE to ASD (S$\to$A). 
From the active speaker detection and audio-visual speech enhancement results, we can conclude that our cyclic unified framework can make the two tasks benefit from each other and significantly improve both speaker detection and speech enhancement performance.

\subsubsection{Real-world Results}
\begin{figure*}[!h]
	\includegraphics[scale=0.45]{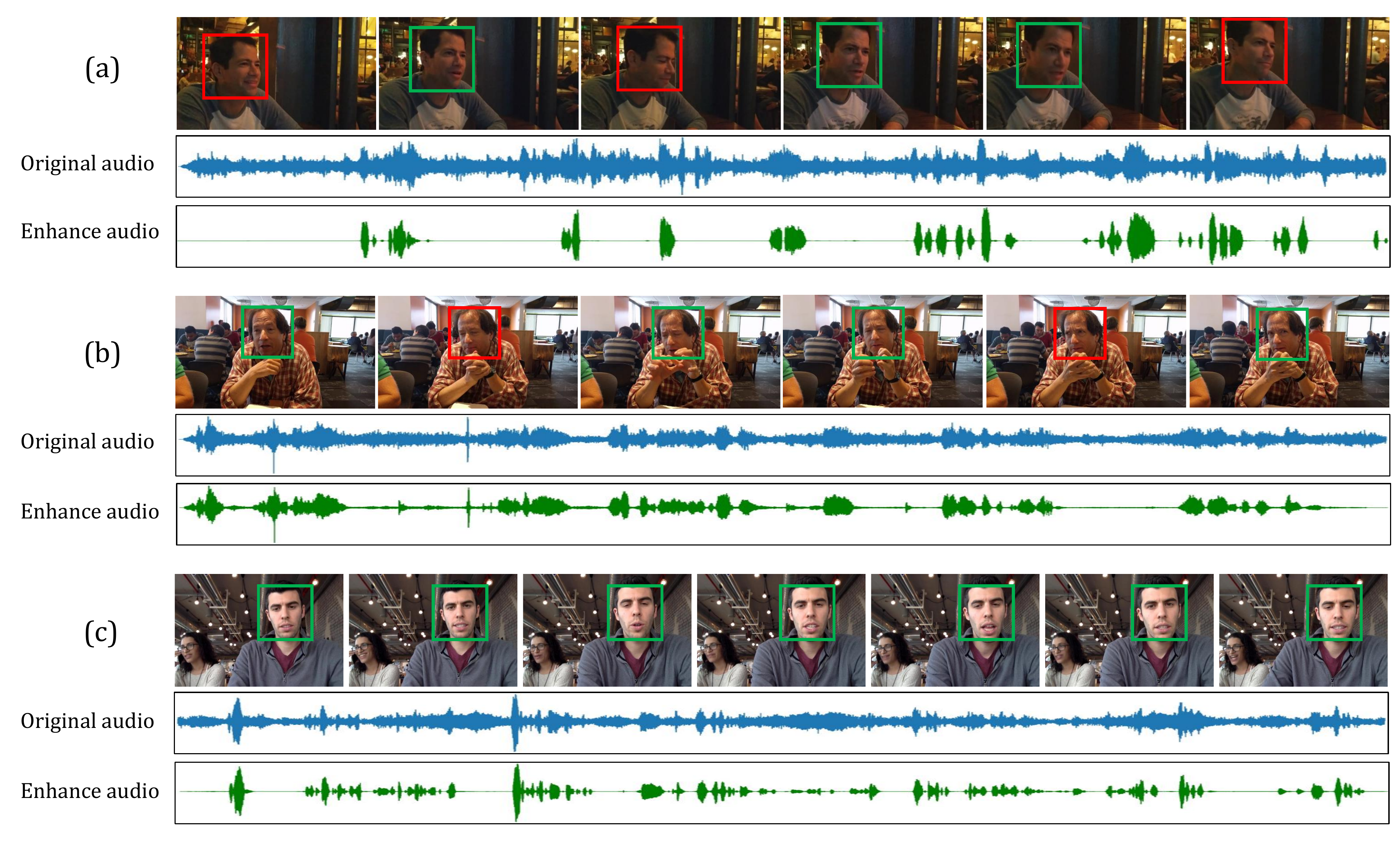}
	\vspace{-10pt}
	\caption{Results of ADENet for the real-world video with speaking in a noise environment. The \textcolor{green}{green box} denotes the speaking speaker, and \textcolor{red}{red box} highlights the non-speaking speaker. (a)The man is speaking in a bar with a loud noise environment. (b)The man is speaking in a restaurant, which includes multiple speakers with low noise. (c)The man is talking into the camera, but there's a woman on the phone next to him. }
	\label{fig:Result}
\end{figure*}

Fig. \ref{fig:Result} presents result samples from real-world videos with one person and multiple persons on the screen. By visualizing active speaker detection and audio-visual speech enhancement achieved by the proposed ADENet framework, it can be easily found that the target speaker has been detected accurately with corresponding enhanced speech signal.

\section{Limitation and Discussion}
The proposed ADENet mainly relies on the temporal relationships of visual and audio modalities to jointly solve the tasks of detection and enhancement. Similar to previous works \cite{tao2017bimodal,tao2019end, tao2021someone} in using temporal network modeling, we introduce conformer block \cite{gulati2020conformer} to extract temporal information from features. However, with the increasing number of conformer, the parameters of the whole model will also rise and lead to more computational cost and longer training time. Therefore, how to optimize the model performance by balancing the parameter scale and computational resources is an issue that needs to be further investigated. A possible way to overcome this limitation is to explore parameter-friendly temporal modeling networks, which may help to find a more efficient solution to a certain extent.

\section{CONCLUSION and Future work}
This work has presented a novel model termed as ADENet for active speaker detection and audio-visual speech enhancement. Key to our method is a unified modeling framework that efficiently learns the different types of multi-modal correlation evidence to generate the aligned audio-visual representations, by introducing cross-modal conformer and simple layer normalization variant.
Thanks to the proposed cross-modal circulant fusion module, our method can effectively detect the target speaker and estimate the speech signal of the target speaker. We demonstrate on the three large-scale benchmark datasets that our model successfully exhibits the advantages of our proposal. The results show that the ADENet significantly outperforms the state-of-the-art.

There are various forms of audio and visual content with different modality compositions in real-world videos. Besides the misleading silent object issue, there exists another challenging case that affects audio-visual learning, as discussed below.

Sound sources are not always visible in video scenarios. For example, the voice of some people chatting may only be background noise in the current scene. Therefore, no corresponding visual object can be used as a condition to isolate the speaker's speech. To be able to solve this problem, we may need to parse the input mixed sound and video frames to identify invisible sounds, and then use other reliable conditions for speech enhancement.

%

\bibliographystyle{IEEEtran}
\bibliography{strings,refs}

\begin{thebibliography}{10}
\providecommand{\url}[1]{#1}
\csname url@samestyle\endcsname
\providecommand{\newblock}{\relax}
\providecommand{\bibinfo}[2]{#2}
\providecommand{\BIBentrySTDinterwordspacing}{\spaceskip=0pt\relax}
\providecommand{\BIBentryALTinterwordstretchfactor}{4}
\providecommand{\BIBentryALTinterwordspacing}{\spaceskip=\fontdimen2\font plus
\BIBentryALTinterwordstretchfactor\fontdimen3\font minus
  \fontdimen4\font\relax}
\providecommand{\BIBforeignlanguage}[2]{{%
\expandafter\ifx\csname l@#1\endcsname\relax
\typeout{** WARNING: IEEEtran.bst: No hyphenation pattern has been}%
\typeout{** loaded for the language `#1'. Using the pattern for}%
\typeout{** the default language instead.}%
\else
\language=\csname l@#1\endcsname
\fi
#2}}
\providecommand{\BIBdecl}{\relax}
\BIBdecl

\bibitem{ERNST2004162}
\BIBentryALTinterwordspacing
M.~O. Ernst and H.~H. Bülthoff, ``Merging the senses into a robust percept,''
  \emph{Trends in Cognitive Sciences}, vol.~8, no.~4, pp. 162--169, 2004.
  [Online]. Available:
  \url{https://www.sciencedirect.com/science/article/pii/S1364661304000385}
\BIBentrySTDinterwordspacing

\bibitem{BURR2006243}
\BIBentryALTinterwordspacing
D.~Burr and D.~Alais, ``Chapter 14 combining visual and auditory information,''
  in \emph{Visual Perception}, ser. Progress in Brain Research,
  S.~Martinez-Conde, S.~Macknik, L.~Martinez, J.-M. Alonso, and P.~Tse,
  Eds.\hskip 1em plus 0.5em minus 0.4em\relax Elsevier, 2006, vol. 155, pp.
  243--258. [Online]. Available:
  \url{https://www.sciencedirect.com/science/article/pii/S0079612306550149}
\BIBentrySTDinterwordspacing

\bibitem{bulthoff1988integration}
H.~H. B{\"u}lthoff and H.~A. Mallot, ``Integration of depth modules: stereo and
  shading,'' \emph{Josa a}, vol.~5, no.~10, pp. 1749--1758, 1988.

\bibitem{alcazar2020active}
J.~L. Alc{\'a}zar, F.~Caba, L.~Mai, F.~Perazzi, J.-Y. Lee, P.~Arbel{\'a}ez, and
  B.~Ghanem, ``Active speakers in context,'' in \emph{Proceedings of the
  IEEE/CVF Conference on Computer Vision and Pattern Recognition}, 2020, pp.
  12\,465--12\,474.

\bibitem{10.1162/jocn.2007.19.12.1964}
\BIBentryALTinterwordspacing
J.~J. Stekelenburg and J.~Vroomen, ``{Neural Correlates of Multisensory
  Integration of Ecologically Valid Audiovisual Events},'' \emph{Journal of
  Cognitive Neuroscience}, vol.~19, no.~12, pp. 1964--1973, 12 2007. [Online].
  Available: \url{https://doi.org/10.1162/jocn.2007.19.12.1964}
\BIBentrySTDinterwordspacing

\bibitem{von2008simulation}
K.~von Kriegstein, {\"O}.~Dogan, M.~Gr{\"u}ter, A.-L. Giraud, C.~A. Kell,
  T.~Gr{\"u}ter, A.~Kleinschmidt, and S.~J. Kiebel, ``Simulation of talking
  faces in the human brain improves auditory speech recognition,''
  \emph{Proceedings of the National Academy of Sciences}, vol. 105, no.~18, pp.
  6747--6752, 2008.

\bibitem{haider2016active}
F.~Haider, N.~Campbell, and S.~Luz, ``Active speaker detection in human machine
  multiparty dialogue using visual prosody information,'' in \emph{2016 IEEE
  global conference on signal and information processing (GlobalSIP)}.\hskip
  1em plus 0.5em minus 0.4em\relax IEEE, 2016, pp. 1207--1211.

\bibitem{roth2020ava}
J.~Roth, S.~Chaudhuri, O.~Klejch, R.~Marvin, A.~Gallagher, L.~Kaver,
  S.~Ramaswamy, A.~Stopczynski, C.~Schmid, Z.~Xi \emph{et~al.}, ``Ava active
  speaker: An audio-visual dataset for active speaker detection,'' in
  \emph{ICASSP 2020-2020 IEEE International Conference on Acoustics, Speech and
  Signal Processing (ICASSP)}.\hskip 1em plus 0.5em minus 0.4em\relax IEEE,
  2020, pp. 4492--4496.

\bibitem{tao2021someone}
R.~Tao, Z.~Pan, R.~K. Das, X.~Qian, M.~Z. Shou, and H.~Li, ``Is someone
  speaking? exploring long-term temporal features for audio-visual active
  speaker detection,'' in \emph{Proceedings of the 29th ACM International
  Conference on Multimedia}, 2021, pp. 3927--3935.

\bibitem{kopuklu2021design}
O.~K{\"o}p{\"u}kl{\"u}, M.~Taseska, and G.~Rigoll, ``How to design a
  three-stage architecture for audio-visual active speaker detection in the
  wild,'' pp. 1193--1203, 2021.

\bibitem{zhang2021unicon}
Y.~Zhang, S.~Liang, S.~Yang, X.~Liu, Z.~Wu, S.~Shan, and X.~Chen, ``Unicon:
  Unified context network for robust active speaker detection,'' in
  \emph{Proceedings of the 29th ACM International Conference on Multimedia},
  2021, pp. 3964--3972.

\bibitem{barnard2014robust}
M.~Barnard, P.~Koniusz, W.~Wang, J.~Kittler, S.~M. Naqvi, and J.~Chambers,
  ``Robust multi-speaker tracking via dictionary learning and identity
  modeling,'' \emph{IEEE Transactions on Multimedia}, vol.~16, no.~3, pp.
  864--880, 2014.

\bibitem{he2016deep}
K.~He, X.~Zhang, S.~Ren, and J.~Sun, ``Deep residual learning for image
  recognition,'' in \emph{Proceedings of the IEEE conference on computer vision
  and pattern recognition}, 2016, pp. 770--778.

\bibitem{tran2015learning}
D.~Tran, L.~Bourdev, R.~Fergus, L.~Torresani, and M.~Paluri, ``Learning
  spatiotemporal features with 3d convolutional networks,'' in
  \emph{Proceedings of the IEEE international conference on computer vision},
  2015, pp. 4489--4497.

\bibitem{chung2019naver}
J.~S. Chung, ``Naver at activitynet challenge 2019--task b active speaker
  detection (ava),'' \emph{arXiv preprint arXiv:1906.10555}, 2019.

\bibitem{zhang2019multi}
Y.-H. Zhang, J.~Xiao, S.~Yang, and S.~Shan, ``Multi-task learning for
  audio-visual active speaker detection,'' \emph{The ActivityNet Large-Scale
  Activity Recognition Challenge}, pp. 1--4, 2019.

\bibitem{li2021ctnet}
Z.~Li, Y.~Sun, L.~Zhang, and J.~Tang, ``Ctnet: Context-based tandem network for
  semantic segmentation,'' \emph{IEEE Transactions on Pattern Analysis and
  Machine Intelligence}, 2021.

\bibitem{gabbay2017visual}
A.~Gabbay, A.~Shamir, and S.~Peleg, ``Visual speech enhancement,'' in
  \emph{Proc. Interspeech}, 2018, pp. 1170--1174.

\bibitem{ephrat2018looking}
A.~Ephrat, I.~Mosseri, O.~Lang, T.~Dekel, K.~Wilson, A.~Hassidim, W.~T.
  Freeman, and M.~Rubinstein, ``Looking to listen at the cocktail party: A
  speaker-independent audio-visual model for speech separation,'' \emph{arXiv
  preprint arXiv:1804.03619}, 2018.

\bibitem{afouras2018deep}
T.~Afouras, J.~S. Chung, A.~Senior, O.~Vinyals, and A.~Zisserman, ``Deep
  audio-visual speech recognition,'' \emph{IEEE transactions on pattern
  analysis and machine intelligence}, 2018.

\bibitem{morrone2019face}
G.~Morrone, S.~Bergamaschi, L.~Pasa, L.~Fadiga, V.~Tikhanoff, and L.~Badino,
  ``Face landmark-based speaker-independent audio-visual speech enhancement in
  multi-talker environments,'' in \emph{ICASSP 2019-2019 IEEE International
  Conference on Acoustics, Speech and Signal Processing (ICASSP)}.\hskip 1em
  plus 0.5em minus 0.4em\relax IEEE, 2019, pp. 6900--6904.

\bibitem{ito2021audio}
K.~Ito, M.~Yamamoto, and K.~Nagamatsu, ``Audio-visual speech enhancement method
  conditioned in the lip motion and speaker-discriminative embeddings,'' in
  \emph{ICASSP 2021-2021 IEEE International Conference on Acoustics, Speech and
  Signal Processing (ICASSP)}.\hskip 1em plus 0.5em minus 0.4em\relax IEEE,
  2021, pp. 6668--6672.

\bibitem{aldeneh2021role}
Z.~Aldeneh, A.~P. Kumar, B.-J. Theobald, E.~Marchi, S.~Kajarekar, D.~Naik, and
  A.~H. Abdelaziz, ``On the role of visual cues in audiovisual speech
  enhancement,'' in \emph{ICASSP 2021-2021 IEEE International Conference on
  Acoustics, Speech and Signal Processing (ICASSP)}.\hskip 1em plus 0.5em minus
  0.4em\relax IEEE, 2021, pp. 8423--8427.

\bibitem{moattar2009simple}
M.~H. Moattar and M.~M. Homayounpour, ``A simple but efficient real-time voice
  activity detection algorithm,'' in \emph{2009 17th European signal processing
  conference}.\hskip 1em plus 0.5em minus 0.4em\relax IEEE, 2009, pp.
  2549--2553.

\bibitem{6737222}
V.~P. Minotto, C.~R. Jung, and B.~Lee, ``Simultaneous-speaker voice activity
  detection and localization using mid-fusion of svm and hmms,'' \emph{IEEE
  Transactions on Multimedia}, vol.~16, no.~4, pp. 1032--1044, 2014.

\bibitem{patrona2016visual}
F.~Patrona, A.~Iosifidis, A.~Tefas, N.~Nikolaidis, and I.~Pitas, ``Visual voice
  activity detection in the wild,'' \emph{IEEE Transactions on Multimedia},
  vol.~18, no.~6, pp. 967--977, 2016.

\bibitem{martin2006enterface}
O.~Martin, I.~Kotsia, B.~Macq, and I.~Pitas, ``The enterface'05 audio-visual
  emotion database,'' in \emph{22nd International Conference on Data
  Engineering Workshops (ICDEW'06)}.\hskip 1em plus 0.5em minus 0.4em\relax
  IEEE, 2006, pp. 8--8.

\bibitem{wu2013two}
C.-H. Wu, J.-C. Lin, and W.-L. Wei, ``Two-level hierarchical alignment for
  semi-coupled hmm-based audiovisual emotion recognition with temporal
  course,'' \emph{IEEE Transactions on Multimedia}, vol.~15, no.~8, pp.
  1880--1895, 2013.

\bibitem{nie2020c}
W.~Nie, M.~Ren, J.~Nie, and S.~Zhao, ``C-gcn: Correlation based graph
  convolutional network for audio-video emotion recognition,'' \emph{IEEE
  Transactions on Multimedia}, vol.~23, pp. 3793--3804, 2020.

\bibitem{neti2000audio}
C.~Neti, G.~Potamianos, J.~Luettin, I.~Matthews, H.~Glotin, D.~Vergyri,
  J.~Sison, and A.~Mashari, ``Audio visual speech recognition,'' IDIAP, Tech.
  Rep., 2000.

\bibitem{tao2020end}
F.~Tao and C.~Busso, ``End-to-end audiovisual speech recognition system with
  multitask learning,'' \emph{IEEE Transactions on Multimedia}, vol.~23, pp.
  1--11, 2020.

\bibitem{liu2020re}
L.~Liu, G.~Feng, D.~Beautemps, and X.-P. Zhang, ``Re-synchronization using the
  hand preceding model for multi-modal fusion in automatic continuous cued
  speech recognition,'' \emph{IEEE Transactions on Multimedia}, vol.~23, pp.
  292--305, 2020.

\bibitem{tao2017bimodal}
F.~Tao and C.~Busso, ``Bimodal recurrent neural network for audiovisual voice
  activity detection.'' in \emph{INTERSPEECH}, 2017, pp. 1938--1942.

\bibitem{tao2019end}
------, ``End-to-end audiovisual speech activity detection with bimodal
  recurrent neural models,'' \emph{Speech Communication}, vol. 113, pp. 25--35,
  2019.

\bibitem{sharma2020crossmodal}
R.~Sharma, K.~Somandepalli, and S.~Narayanan, ``Crossmodal learning for
  audio-visual speech event localization,'' \emph{arXiv preprint
  arXiv:2003.04358}, 2020.

\bibitem{shvets2019leveraging}
M.~Shvets, W.~Liu, and A.~C. Berg, ``Leveraging long-range temporal
  relationships between proposals for video object detection,'' in
  \emph{Proceedings of the IEEE/CVF International Conference on Computer
  Vision}, 2019, pp. 9756--9764.

\bibitem{golumbic2013visual}
E.~Z. Golumbic, G.~B. Cogan, C.~E. Schroeder, and D.~Poeppel, ``Visual input
  enhances selective speech envelope tracking in auditory cortex at a
  “cocktail party”,'' \emph{Journal of Neuroscience}, vol.~33, no.~4, pp.
  1417--1426, 2013.

\bibitem{afouras2018conversation}
T.~Afouras, J.~S. Chung, and A.~Zisserman, ``The conversation: Deep
  audio-visual speech enhancement,'' in \emph{Interspeech}, 2018.

\bibitem{afouras2020self}
T.~Afouras, A.~Owens, J.~S. Chung, and A.~Zisserman, ``Self-supervised learning
  of audio-visual objects from video,'' in \emph{Computer Vision--ECCV 2020:
  16th European Conference, Glasgow, UK, August 23--28, 2020, Proceedings, Part
  XVIII 16}.\hskip 1em plus 0.5em minus 0.4em\relax Springer, 2020, pp.
  208--224.

\bibitem{owens2018audio}
A.~Owens and A.~A. Efros, ``Audio-visual scene analysis with self-supervised
  multisensory features,'' in \emph{Proceedings of the European Conference on
  Computer Vision (ECCV)}, 2018, pp. 631--648.

\bibitem{gao2021visualvoice}
R.~Gao and K.~Grauman, ``Visualvoice: Audio-visual speech separation with
  cross-modal consistency,'' in \emph{2021 IEEE/CVF Conference on Computer
  Vision and Pattern Recognition (CVPR)}.\hskip 1em plus 0.5em minus
  0.4em\relax IEEE, 2021, pp. 15\,490--15\,500.

\bibitem{Morrone2019}
G.~Morrone, S.~Bergamaschi, L.~Pasa, L.~Fadiga, V.~Tikhanoff, and L.~Badino,
  ``{Face Landmark-based Speaker-independent Audio-visual Speech Enhancement in
  Multi-talker Environments},'' \emph{ICASSP, IEEE International Conference on
  Acoustics, Speech and Signal Processing - Proceedings}, vol. 2019-May, pp.
  6900--6904, 2019.

\bibitem{KING2005R339}
\BIBentryALTinterwordspacing
A.~J. King, ``Multisensory integration: Strategies for synchronization,''
  \emph{Current Biology}, vol.~15, no.~9, pp. R339--R341, 2005. [Online].
  Available:
  \url{https://www.sciencedirect.com/science/article/pii/S0960982205004227}
\BIBentrySTDinterwordspacing

\bibitem{8403294}
Z.~Li, J.~Tang, and T.~Mei, ``Deep collaborative embedding for social image
  understanding,'' \emph{IEEE Transactions on Pattern Analysis and Machine
  Intelligence}, vol.~41, no.~9, pp. 2070--2083, 2019.

\bibitem{li2016weakly}
Z.~Li and J.~Tang, ``Weakly supervised deep matrix factorization for social
  image understanding,'' \emph{IEEE Transactions on Image Processing}, vol.~26,
  no.~1, pp. 276--288, 2016.

\bibitem{li2016multimedia}
Z.~Li, J.~Tang, X.~Wang, J.~Liu, and H.~Lu, ``Multimedia news summarization in
  search,'' \emph{ACM Transactions on Intelligent Systems and Technology
  (TIST)}, vol.~7, no.~3, pp. 1--20, 2016.

\bibitem{gao2020listen}
R.~Gao, T.-H. Oh, K.~Grauman, and L.~Torresani, ``Listen to look: Action
  recognition by previewing audio,'' in \emph{Proceedings of the IEEE/CVF
  Conference on Computer Vision and Pattern Recognition}, 2020, pp.
  10\,457--10\,467.

\bibitem{kazakos2019epic}
E.~Kazakos, A.~Nagrani, A.~Zisserman, and D.~Damen, ``Epic-fusion: Audio-visual
  temporal binding for egocentric action recognition,'' in \emph{Proceedings of
  the IEEE/CVF International Conference on Computer Vision}, 2019, pp.
  5492--5501.

\bibitem{hu2021class}
D.~Hu, Y.~Wei, R.~Qian, W.~Lin, R.~Song, and J.-R. Wen, ``Class-aware sounding
  objects localization via audiovisual correspondence,'' \emph{IEEE
  Transactions on Pattern Analysis and Machine Intelligence}, 2021.

\bibitem{rao2021decompose}
V.~R. Rao, M.~I. Khalil, H.~Li, P.~Dai, and J.~Lu, ``Decompose the sounds and
  pixels, recompose the events,'' \emph{arXiv preprint arXiv:2112.11547}, 2021.

\bibitem{zhou2021pose}
H.~Zhou, Y.~Sun, W.~Wu, C.~C. Loy, X.~Wang, and Z.~Liu, ``Pose-controllable
  talking face generation by implicitly modularized audio-visual
  representation,'' in \emph{Proceedings of the IEEE/CVF Conference on Computer
  Vision and Pattern Recognition}, 2021, pp. 4176--4186.

\bibitem{wang2021one}
S.~Wang, L.~Li, Y.~Ding, and X.~Yu, ``One-shot talking face generation from
  single-speaker audio-visual correlation learning,'' \emph{arXiv preprint
  arXiv:2112.02749}, 2021.

\bibitem{hu2018squeeze}
J.~Hu, L.~Shen, and G.~Sun, ``Squeeze-and-excitation networks,'' in
  \emph{Proceedings of the IEEE conference on computer vision and pattern
  recognition}, 2018, pp. 7132--7141.

\bibitem{gulati2020conformer}
A.~Gulati, J.~Qin, C.-C. Chiu, N.~Parmar, Y.~Zhang, J.~Yu, W.~Han, S.~Wang,
  Z.~Zhang, Y.~Wu \emph{et~al.}, ``Conformer: Convolution-augmented transformer
  for speech recognition,'' in \emph{Proc. Interspeech}, 2020.

\bibitem{zhangictcas}
Y.~Zhang, S.~Liang, S.~Yang, X.~Liu, Z.~Wu, and S.~Shan, ``Ictcas-ucas-tal
  submission to the ava-activespeaker task at activitynet challenge 2021.''

\bibitem{van2008visualizing}
L.~Van~der Maaten and G.~Hinton, ``Visualizing data using t-sne.''
  \emph{Journal of machine learning research}, vol.~9, no.~11, 2008.

\bibitem{ba2016layer}
J.~L. Ba, J.~R. Kiros, and G.~E. Hinton, ``Layer normalization,'' \emph{arXiv
  preprint arXiv:1607.06450}, 2016.

\bibitem{michelsanti2021overview}
D.~Michelsanti, Z.-H. Tan, S.-X. Zhang, Y.~Xu, M.~Yu, D.~Yu, and J.~Jensen,
  ``An overview of deep-learning-based audio-visual speech enhancement and
  separation,'' \emph{IEEE/ACM Transactions on Audio, Speech, and Language
  Processing}, 2021.

\bibitem{oord2016wavenet}
A.~v.~d. Oord, S.~Dieleman, H.~Zen, K.~Simonyan, O.~Vinyals, A.~Graves,
  N.~Kalchbrenner, A.~Senior, and K.~Kavukcuoglu, ``Wavenet: A generative model
  for raw audio,'' \emph{arXiv preprint arXiv:1609.03499}, 2016.

\bibitem{luo2019conv}
Y.~Luo and N.~Mesgarani, ``Conv-tasnet: Surpassing ideal time--frequency
  magnitude masking for speech separation,'' \emph{IEEE/ACM transactions on
  audio, speech, and language processing}, vol.~27, no.~8, pp. 1256--1266,
  2019.

\bibitem{chung2018voxceleb2}
J.~S. Chung, A.~Nagrani, and A.~Zisserman, ``Voxceleb2: Deep speaker
  recognition,'' in \emph{INTERSPEECH}, 2018.

\bibitem{afouras2018lrs3}
T.~Afouras, J.~S. Chung, and A.~Zisserman, ``Lrs3-ted: a large-scale dataset
  for visual speech recognition,'' \emph{arXiv preprint arXiv:1809.00496},
  2018.

\bibitem{chakravarty2016cross}
P.~Chakravarty and T.~Tuytelaars, ``Cross-modal supervision for learning active
  speaker detection in video,'' in \emph{European Conference on Computer
  Vision}.\hskip 1em plus 0.5em minus 0.4em\relax Springer, 2016, pp. 285--301.

\bibitem{snyder2015musan}
D.~Snyder, G.~Chen, and D.~Povey, ``Musan: A music, speech, and noise corpus,''
  \emph{arXiv preprint arXiv:1510.08484}, 2015.

\bibitem{lienhart2001reliable}
R.~Lienhart, ``Reliable transition detection in videos: A survey and
  practitioner's guide,'' \emph{International journal of image and graphics},
  vol.~1, no.~03, pp. 469--486, 2001.

\bibitem{rix2001perceptual}
A.~W. Rix, J.~G. Beerends, M.~P. Hollier, and A.~P. Hekstra, ``Perceptual
  evaluation of speech quality (pesq)-a new method for speech quality
  assessment of telephone networks and codecs,'' in \emph{2001 IEEE
  international conference on acoustics, speech, and signal processing.
  Proceedings (Cat. No. 01CH37221)}, vol.~2.\hskip 1em plus 0.5em minus
  0.4em\relax IEEE, 2001, pp. 749--752.

\bibitem{kingma2014adam}
D.~P. Kingma and J.~Ba, ``Adam: A method for stochastic optimization,'' in
  \emph{ICLR (Poster)}, 2015.

\bibitem{leon2021maas}
J.~L. Alc{\'a}zar, F.~Caba, A.~K. Thabet, and B.~Ghanem, ``Maas: Multi-modal
  assignation for active speaker detection,'' in \emph{Proceedings of the
  IEEE/CVF International Conference on Computer Vision}, 2021, pp. 265--274.

\bibitem{9580127}
R.~Ray, S.~Karthik, V.~Mathur, P.~Kumar, M.~G, S.~Tiwari, and R.~T.
  Shankarappa, ``Feature genuinization based residual squeeze-and-excitation
  for audio anti-spoofing in sound ai,'' in \emph{2021 12th International
  Conference on Computing Communication and Networking Technologies (ICCCNT)},
  2021, pp. 1--5.

\bibitem{zach2007duality}
C.~Zach, T.~Pock, and H.~Bischof, ``A duality based approach for realtime tv-l
  1 optical flow,'' in \emph{Joint pattern recognition symposium}.\hskip 1em
  plus 0.5em minus 0.4em\relax Springer, 2007, pp. 214--223.

\bibitem{chung2016out}
J.~S. Chung and A.~Zisserman, ``Out of time: automated lip sync in the wild,''
  in \emph{Asian conference on computer vision}.\hskip 1em plus 0.5em minus
  0.4em\relax Springer, 2016, pp. 251--263.

\bibitem{shahid2021s}
M.~Shahid, C.~Beyan, and V.~Murino, ``S-vvad: Visual voice activity detection
  by motion segmentation,'' in \emph{Proceedings of the IEEE/CVF Winter
  Conference on Applications of Computer Vision}, 2021, pp. 2332--2341.

\bibitem{truong2021right}
T.-D. Truong, C.~N. Duong, H.~A. Pham, B.~Raj, N.~Le, K.~Luu \emph{et~al.},
  ``The right to talk: An audio-visual transformer approach,'' in
  \emph{Proceedings of the IEEE/CVF International Conference on Computer
  Vision}, 2021, pp. 1105--1114.

\bibitem{hershey2016deep}
J.~R. Hershey, Z.~Chen, J.~Le~Roux, and S.~Watanabe, ``Deep clustering:
  Discriminative embeddings for segmentation and separation,'' in \emph{2016
  IEEE International Conference on Acoustics, Speech and Signal Processing
  (ICASSP)}.\hskip 1em plus 0.5em minus 0.4em\relax IEEE, 2016, pp. 31--35.

\end{thebibliography}
 
\newpage
\vfill

\end{document}